\documentclass[12pt]{article}
\usepackage{enumerate}
\usepackage[
  authordate,
  backend=biber,
  natbib=true,
  giveninits=true,
  maxcitenames=1,
  mincitenames=1,
  doi=false,
    url=false,
    isbn=false,
    eprint=false,
  uniquelist=false,
  uniquename=false
]{biblatex-chicago}

\usepackage{xr} 
\usepackage{caption}
\usepackage{subcaption}
\usepackage{url} % not crucial - just used below for the URL 
\usepackage{amsthm,amsmath,amssymb,bm}
\usepackage{xcolor} 
\definecolor{myblue}{RGB}{25, 120, 200}
\definecolor{mygreen}{RGB}{32, 178, 120}
\usepackage[colorlinks=true, linkcolor=myblue, citecolor=mygreen]{hyperref} % 使用自定义颜色
\AtEveryCitekey{\color{mygreen}}
\usepackage{algorithm}
\usepackage{algpseudocode}

\usepackage{graphicx}
\usepackage{mathabx}
\usepackage{comment}
\usepackage{xr} 
\newcommand{\blind}{0}

\newcommand{\E}{\mathbb{E}}
\newcommand{\Var}{\mathrm{Var}}

\newcommand{\Cov}{\mathrm{Cov}}

\newcommand{\W}{\mathcal{W}}
\newcommand{\Tcal}{\mathcal{T}}
\newcommand{\Xcal}{\mathcal{X}}
\newcommand{\Ical}{\mathcal{I}}

\usepackage{amsthm}

\theoremstyle{plain}
\newtheorem{theorem}{Theorem}
\newtheorem{lemma}{Lemma}
\newtheorem{proposition}{Proposition}
\newtheorem{corollary}{Corollary}

\theoremstyle{definition}
\newtheorem{assumption}{Assumption}
\newtheorem{example}{Example}

\theoremstyle{remark}
\newtheorem{remark}{Remark}

\begin{document}

\def\spacingset#1{\renewcommand{\baselinestretch}%
{#1}\small\normalsize} \spacingset{1}

\newcommand{\ignore}[1]{}

%%%%%%%%%%%%%%%%%%%%%%%%%%%%%%%%%%%%%%%%%%%%%%%%%%%%%%%%%%%%%%%%%%%%%%%%%%%%%%

\if1\blind
{
  \title{\bf  Generation-Powered Inference for Distribution-Valued Outcomes}
      \date{}
  \maketitle
} \fi

\if0\blind
{
  \bigskip
  \bigskip
  \bigskip
\begin{center}
    \setlength{\baselineskip}{1.5\baselineskip}
    {\LARGE\bf      Generation-Powered Inference for Distribution-Valued Outcomes}\\
    \medskip\medskip\medskip
   {\large Yijiao Zhang and Hongzhe Li\\
   Department of Biostatistics, Epidemiology and Informatics\\
   University of Pennsylvania
    }
\end{center}
  \medskip
} \fi

\begin{abstract}
Modern generative models increasingly produce distribution-valued outputs, such as predicted cellular responses to genetic perturbations in single-cell genomics. While these models provide valuable auxiliary information, they are inherently imperfect, creating a need for statistical methods that leverage their predictions without relying on their correctness. We propose generation-powered inference (GPI), a general framework for improving inference on distribution-valued parameters using auxiliary generative models.
Focusing on Wasserstein barycenters and related distributional functionals, we introduce a function-valued bridge representation that transforms inference in the nonlinear Wasserstein space into estimation of a mean function in a Hilbert space, enabling an augmented estimation framework analogous to prediction-powered inference. We develop a family of GPI estimators with optimal information borrowing, establish consistency, asymptotic normality, and simultaneous confidence bands, and derive valid inference for linear functionals and Wasserstein distances.
Simulation studies demonstrate efficiency gains over labeled-data-only methods and robust performance under generative model misspecification. We illustrate the proposed framework using a Perturb-seq study of K562 cells, where synthetic perturbation responses generated by the State foundation model are used to improve inference for pathway-level consensus gene expression distributions associated with perturbations of the 40S ribosome module.
\end{abstract}


\spacingset{1.8} % DON'T change the spacing!

\section{Introduction}

Modern generative models increasingly produce synthetic outcomes that can be viewed as draws from an estimated distribution.  Generative foundation models for single-cell genomics generate distributions of cellular states under unseen perturbations \citep{adduri2025predicting}; autoregressive large language models generate distributions over text continuations, allowing plausible responses to be sampled from the same prompt \citep{brown2020language}; climate models generate predictive distributions over future trajectories \citep{Price2025}; and digital twin technologies generate synthetic representations of individuals or populations for simulation, forecasting, and scientific and clinical decision making \citep{Sharma2022DigitalTwins}. As these generative models become increasingly integrated into scientific workflows, they offer the potential to substantially augment limited experimental or observational data. However, despite their practical utility, generative models are inherently imperfect \citep{farquhar2024detecting,ahlmann2025deep}. They are trained on finite and potentially biased datasets, rely on simplifying assumptions, and may fail to generalize to new domains. This raises a fundamental statistical question: how can one leverage AI-generated distributions to improve statistical efficiency while maintaining valid inference in the presence of model misspecification?

Recent work on prediction-powered inference (PPI) and related methods has demonstrated that auxiliary predictions from machine learning models can substantially improve statistical inference for finite-dimensional parameters while remaining robust to prediction errors \citep{Angelopoulos2023PPI,Angelopoulos2023PPIpp}. These methods have established a general paradigm in which predictions are used to augment limited labeled data through bias-corrected estimating equations, leading to valid and often more efficient estimators. However, existing prediction-powered methods have primarily focused on finite-dimensional Euclidean parameters \citep{ji2025predictions,xu2025unified,miao2025assumption}, including means and regression coefficients. They do not address inference when the object of interest is itself distribution-valued.

Distribution-valued parameters arise naturally in a wide range of modern applications. In biomedicine, perturbation experiments seek to characterize how genetic or chemical interventions reshape the distribution of cellular states, rather than merely their average effects \citep{song2025decoding}.  In population health, one may wish to compare distributions of daily activity profiles across populations \citep{lin2022causal}. In economics and social sciences, policy interventions are often evaluated through their effects on income or outcome distributions \citep{bitler2006mean}. More broadly, scientific questions increasingly concern how entire distributions change across conditions, motivating inference on objects such as Wasserstein barycenters, quantile functions, transport maps, and other nonlinear distributional summaries.

Statistical inference for distribution-valued outcomes presents unique challenges. Unlike Euclidean outcomes,  distributions equipped with the Wasserstein metric form a nonlinear geometric space \citep{panaretos2020invitation,villani2009optimal}. Consequently, many classical semiparametric tools, including additive estimating equations and bias-correction, do not apply directly. This difficulty is particularly pronounced when the available observations are sparse or incomplete and generative models are introduced to compensate for missing information. In such settings, naively treating generated samples as if they were observed data may lead to invalid uncertainty quantification and misleading scientific conclusions.

In this paper, we develop a general framework for generation-powered inference (GPI) for distribution-valued parameters. Our central idea is to introduce a bridge representation that transforms inference in Wasserstein space into estimation of a mean element in a Hilbert space. Specifically, for a Wasserstein barycenter, we define a function-valued bridge parameter by composing its quantile function with a reference distribution. For a given barycenter distribution function $\mu_\oplus$, the function-valued target bridge parameter is defined as $$\theta^\lambda(\cdot)
:=\mu_\oplus^{-1}\circ\lambda(\cdot),$$ where $\mu_\oplus^{-1}$ is the quantile function of the distribution $\mu_{\oplus}$ and $\lambda$ is a continuous reference distribution function. A crucial component in this definition is the choice of reference distribution $\lambda$, that is allowed to
have a domain different from that of $\mu_\oplus$.  This representation linearizes a broad class of nonlinear distributional inferential problems and enables principled augmentation with AI-generated distributional data. The resulting framework extends prediction-powered inference from finite-dimensional Euclidean settings to nonlinear distribution-valued targets.

The bridge representation serves as a unifying object for downstream inference. Depending on the choice of reference distribution, it encodes the quantile function of the barycenter, optimal transport maps, displacement functions, Wasserstein distances, and a broad class of linear and nonlinear distributional summaries. Building on this representation, we develop a family of generation-powered estimators that combine experimentally observed and AI-generated distributions through an operator-valued augmentation mechanism. We establish consistency, asymptotic normality, simultaneous confidence bands, and inference procedures for a variety of downstream functionals. Furthermore, we characterize the optimal information-borrowing operator and show that the resulting estimator is asymptotically optimal within a broad class of augmentation procedures.

Our work is motivated by applications in single-cell perturbation genomics \citep{Dixit2016PerturbSeq,replogle2022mapping,song2025decoding}, where recent generative models, including emerging single-cell foundation models, can generate cellular responses to genetic perturbations that have not been experimentally assayed \citep{adduri2025predicting, klein2025cellflow, yuan2026perturbdiff, he2026squidiff}. In large-scale Perturb-seq studies, investigators are often interested in inferring the consensus effect of biological pathways or gene modules, yet perturbation measurements are frequently sparse or entirely missing for many genes in a biological pathway or module \citep{khoroshkin2024posttranscriptional,jiang2025mixscale}. Generation-powered inference provides a principled approach for integrating observed and synthetic cellular responses while accounting for uncertainty arising from both finite samples and imperfect generative models. More broadly, the proposed framework is applicable whenever generative models are used to augment data for inference on distribution-valued quantities.

The main contributions of this paper are fourfold. First, we introduce a bridge representation that converts inference for Wasserstein barycenters into estimation of a mean function in a Hilbert space. Second, we develop a generation-powered inference framework that leverages auxiliary generative models while remaining robust to model misspecification. Third, we establish a comprehensive asymptotic theory covering Gaussian process limits, feasible and valid downstream inference, and efficiency guarantees for the optimal operator-valued augmentation. Finally, we demonstrate the practical utility of the proposed methods through simulations and an analysis of a large Perturb-seq study using the State single-cell foundation model.

The remainder of the paper is organized as follows. Section \ref{sec:PPI-and-W2} reviews prediction-powered inference and introduces distribution-valued outcomes.  Section \ref{sec:GPI} presents the bridge parameter construction and the proposed generation-powered estimators.  Section \ref{sec:asymptotic} develops the asymptotic theory and optimal information-borrowing results.  Section \ref{sec:inference} presents inference procedures for bridge parameters and their associated distributional functionals. Sections \ref{sec:simu} and \ref{sec:real} present simulation studies and a real-data application.   Section \ref{sec:Discuss} provides a brief discussion. 
%Technical details and proofs are provided in the Supplementary Material.

\section{Prediction-powered Inference  and Distribution-valued Outcomes}\label{sec:PPI-and-W2}

\subsection{Prediction-powered inference as an augmented estimator}\label{subsec:PPI-and-W2}

Let \(X\in\mathbb R^d\) denotes the covariates and \(Y\in\mathbb R\) denotes the outcome of interest. Suppose we observe a small labeled sample $\{(X_i,Y_i)\}_{i\in\mathcal O}$ with size \(n=|\mathcal O|\)
and a large unlabeled dataset $\{X_j\}_{j\in\mathcal U}$ with size \(N=|\mathcal U|\). The target parameter \(\theta\) is defined by the population risk minimization problem
$\theta
=
\arg\min_\theta
\E\{\ell_\theta(X,Y)\},$
where \(\ell_\theta\) is a loss function, or equivalently by the estimating equation $\E\{\nabla \ell_{\theta}(X,Y)\}=0,$ where \(\nabla \ell_{\theta}:\mathcal X\times\mathcal Y\to\mathbb R^p\) is a subgradient of \(\ell_\theta\) with respect to \(\theta\). An estimator based only on the labeled sample is limited in efficiency by the sample size \(n\).

Now suppose that, for a given covariate \(X\), a prediction model \(f(X)\) is available, which may be possibly misspecified. Prediction-powered inference aims to leverage such auxiliary predictions \(f(X)\) to improve the efficiency of estimating \(\theta\) \citep{Angelopoulos2023PPI}. Specifically, the PPI estimator is defined as
\[
\hat\theta_{\mathrm{PPI}}
=
\arg\min_\theta
\left[
\frac{1}{N}\sum_{j\in\mathcal U}
\ell_\theta\bigl(X_j,f(X_j)\bigr)
+
\frac{1}{n}\sum_{i\in\mathcal O}
\Bigl\{
\ell_\theta(X_i,Y_i)-\ell_\theta\bigl(X_i,f(X_i)\bigr)
\Bigr\}
\right].
\]

A useful perspective is to interpret PPI through the lens of missing data. Let \(S_i=1\) indicate that \(Y_i\) is observed, that is, \(i\in\mathcal O\), and let \(S_i=0\) indicate that \(i\in\mathcal U\). If we are willing to assume that the outcome is missing completely at random (MCAR), namely $S\perp (X,Y), \mathbb P(S=1)=\pi\in(0,1),$
then \(\theta\) can equivalently be identified through the augmented estimating equation
\begin{equation}
\label{eq:aug_eq_scalar}
\E\left[
\frac{S}{\pi}\{\nabla \ell_{\theta}(X,Y)-m_\theta(X)\}
+m_\theta(X)
\right]
=0,
\end{equation}
where \(m_\theta(X)\) is a working surrogate for \(\nabla \ell_{\theta}(X,Y)\). This equation remains unbiased for any $m_{\theta}$, and attains the semiparametric efficiency bound when $m_{\theta}$ is correctly specified \citep{tsiatis_semiparametric_2007}. 

The PPI estimator \citep{Angelopoulos2023PPI,Zrnic2024CrossPPI} corresponds to choosing $m_\theta(X)=\nabla \ell_{\theta}\bigl(X,f(X)\bigr)$ in \eqref{eq:aug_eq_scalar} and therefore can be viewed as a particular augmented estimating equation that uses the predictor \(f(X)\) to construct a working surrogate for the complete-data estimating function. More generally, an appropriate choice of \(m_\theta\) can ensure that the resulting estimator is at least as efficient as its labeled-only counterpart \citep{Angelopoulos2023PPIpp}.

The population mean of the discrepancy term $
\nabla \ell_{\theta}(X,Y)-m_\theta(X)
$ in (\ref{eq:aug_eq_scalar}) is referred to as the \emph{rectifier} in the PPI literature, which captures how prediction errors bias the  the gold-standard subgradient of the loss function $\mathbb{E}[\nabla \ell_{\theta}\bigl(X,Y\bigr)]$.

\subsection{Distribution-valued outcomes and Wasserstein space}\label{subsec:W2}
%Perturbations in heterogeneous cell populations often lead to shifts in the mean state, changes in variance, changes in subpopulation proportions, and appearance or disappearance of cell subtypes. We therefore characterize perturbation responses by comparing the full distributions of cellular outcomes.

Let \(\mathcal I\subset\mathbb R\) be an interval. For two distribution functions \(\lambda_1\) and \(\lambda_2\) on \(\mathcal I\) with finite second moments, let $\Lambda(\lambda_1,\lambda_2)$
denote the set of all joint distributions on \(\mathcal I\times\mathcal I\) having marginals \(\lambda_1\) and \(\lambda_2\). The 2-Wasserstein ($W_2$) distance is defined as
\[
W_2(\lambda_1,\lambda_2)
:=
\left(
\inf_{\lambda_{12} \in\Lambda(\lambda_1,\lambda_2)}
\int_{\mathcal I\times\mathcal I}
(s-t)^2\,d\lambda_{12}(s,t)
\right)^{1/2}.
\]

The Wasserstein distance measures the minimum cost of transporting one distribution into another and thereby respects the geometry of the sample space. This perspective motivated optimal transport methods for modeling heterogeneous single-cell perturbation responses \citep{bunne2023learning,dong2023causal}. 

The Wasserstein space of order 2 on \(\mathcal I\) is defined as $$\mathcal W_2(\mathcal I)
:=\{
\mu \text{ is a distribution function on } \mathcal I:
\int_{\mathcal I} t^2\,d\mu(t)<\infty
\},$$
endowed with the $W_2$ distance. For any distribution function
\(\lambda\) on \(\mathcal I\), we define its generalized quantile function by $\mu^{-1}(u)
:=
\inf\{t\in\mathcal I:\mu(t)\ge u\}, u\in(0,1).$ In one dimension, $W_2$ admits the following quantile representation, which underlies our bridge construction.
\begin{proposition}
\label{prop:w2-representation}
Let \(\mu,\nu\in\mathcal W_2(\mathcal I)\), and \(\lambda\) be a continuous distribution function. Then $W_2^2(\mu,\nu)
=
\int
\bigl\{
\mu^{-1}\circ\lambda(t)-\nu^{-1}\circ\lambda(t)
\bigr\}^2\,d\lambda(t).$
\end{proposition}

%The Wasserstein metric is particularly appealing for distribution-valued data because it respects the geometry of the sample space. It is also well suited for heterogeneous cell populations resulted from Perturb-seq perturbations, which often lead to  shifts in the mean state, changes in variance, changes in subpopulation proportions, and appearance or disappearance of cell subtypes. Wasserstein distance naturally captures all these effects in a unified way as  it compares the entire distribution.

We extend the setup of Section~\ref{subsec:PPI-and-W2} to a distribution-valued outcome $Y\in\mathcal W_2(\mathcal I)$, replacing $f(x)$ with a generative surrogate $G(x)\in\mathcal W_2(\mathcal I)$.  For each covariate value \(x\in\Xcal\), \(G(x)\) represents a surrogate distribution from which synthetic samples can be generated. Ideally, \(G(X)\) serves as an informative distribution-valued surrogate for \(Y\), but we do not require the two to coincide, allowing the generative model to be systematically imperfect. A natural target parameter is the Wasserstein barycenter of $Y$, defined by $\mu_\oplus:=\arg\min_{\mu\in\mathcal W_2(\mathcal I)}\E\{
W_2^2(Y,\mu)\}.$

%A direct extension of PPI to distribution-valued outcomes would be to define a loss function $\ell_\theta(X,Y):\mathcal X\times \mathcal W_2(\mathcal I)\to\mathbb R,$ for example through a Wasserstein loss, and then construct the rectifier using the corresponding ``gradient'' difference. While such an approach is possible in principle, it faces several difficulties. First, the classical PPI estimator is essentially a linear bias-correction estimator. The key reason is that the gradient of the loss takes values in a linear space, so an additive rectifier can be defined naturally. In contrast, the parameter of interest here, such as the Wasserstein barycenter \(\mu_\oplus\), lies in \(\mathcal W_2(\mathcal I)\), which is a nonlinear space. Consequently, an additive rectifier can no longer be introduced in the same direct manner. Second, a key advantage of classical PPI is that its asymptotic variance can be characterized explicitly, which allows one to construct valid confidence intervals and related inferential procedures. When the target parameter itself is distribution-valued, however, the corresponding gradient is no longer an ordinary Euclidean gradient, and the resulting asymptotic variance becomes much less tractable.
 
A direct extension of PPI to distribution-valued outcomes would be to define a loss function $\ell_\theta(X,Y):\mathcal X\times \mathcal W_2(\mathcal I)\to\mathbb R,$ for example through a Wasserstein loss, and construct the rectifier using the corresponding ``gradient'' discrepancy. Although possible in principle, this approach faces two main difficulties. First, PPI relies on additive bias correction because the gradient of the loss takes values in a linear space. In contrast, a distribution-valued target such as the Wasserstein barycenter \(\mu_\oplus\) lies in the nonlinear space \(\mathcal W_2(\mathcal I)\), where an additive rectifier is not directly defined. Second, the non-Euclidean nature of the target complicates the characterization of its asymptotic distribution and the construction of valid inferential procedures.

These observations suggest that the central challenge is not to extend the PPI machinery directly to the nonlinear Wasserstein space, but rather to identify a suitable linear representation of the distribution-valued target that supports additive augmentation. A fundamental property of one-dimensional Wasserstein geometry is that the barycenter admits such a representation through its quantile function \citep{villani2003topics}, which motivates the construction developed in the next section.

%However, the discussion in Section \ref{subsec:PPI-and-W2} motivates us to revisit the problem from the viewpoint of missing data. Thanks to several pioneering works in causal inference for distribution-valued outcomes \citep{lin2022causal}, it becomes possible to define the rectifier in a more geometrically meaningful and statistically tractable way. 

\section{Generation-Powered Inference for Distributional Outcomes}\label{sec:GPI}
Building on the distribution-valued setup introduced in Section \ref{subsec:W2}, we now develop our generation-powered inference framework. 
%We adopt the basic setup introduced in Section \ref{subsec:PPI-and-W2}, except that the outcome $Y_i$ now is a distribution function and resides in $\mathcal W_2(\mathcal I).$ Let $G:\mathcal X\to\mathcal W_2(\mathcal I)$ denote a conditional generative model for $Y$ given $X$, which maps each covariate value  \(x\) to a synthetic outcome distribution \(G(x)\).

\subsection{A bridge parameter for distributional inference}
In many applications the primary interest is not the barycenter as an abstract distribution object, but rather a collection of downstream tasks built upon it. Examples include inference on the quantile function of \(\mu_\oplus\), sampling from an estimated barycenter distribution, and comparing an estimated barycenter with a given control distribution. Therefore, motivated by the discussion in Section \ref{sec:PPI-and-W2} , we seek a function-valued bridge parameter satisfying the following two properties:

\begin{enumerate}

    \item[(a)] It should admit an asymptotically linear representation, so that a proper rectifier can be naturally defined.

    \item[(b)] It should embed the information needed for downstream tasks.
\end{enumerate}

To formalize the idea, we introduce \(\lambda\) as a continuous reference distribution function with domain $\Tcal$. We define the function-valued target parameter as
\begin{equation}
\label{eq:target_theta}
\theta^\lambda(\cdot)
:=\mu_\oplus^{-1}\circ\lambda(\cdot).
\end{equation}
This transformation embeds the distribution-valued outcome into a common linear function space indexed by the reference distribution \(\lambda\). We claim that $\theta^\lambda$ is a proper bridge parameter that satisfies the two properties listed above. The following lemma shows that $\theta^\lambda$ enjoys a linear representation.
 
\begin{lemma}
\label{lem:barycenter_transform}  Let $\mu_\oplus$ be the barrycenter of distribution $Y\in \mathcal W_2(\mathcal I)$.
For a distribution function $\lambda$, $\mu_\oplus^{-1}\circ\lambda(\cdot)=\E\!\left[
Y^{-1}\circ\lambda(\cdot)
\right]$.
\end{lemma}
Therefore, by working with the transformed outcome, inference based on the barycenter is reduced to estimation of an ordinary mean function. Throughout the paper, we assume that the reference distribution \(\lambda\) is known. The bridge parameter is a canonical representation of the barycenter, while different choices of the reference distribution $\lambda$ adapt this representation to different inferential goals. 

For a region \(A=[p,q]\subset[0,1]\), define the weight function $\ell_A^\lambda(t)
=(q-p)^{-1}\mathbf 1\{\lambda(t)\in A\}$. For a reference distribution function \(\lambda\), define $L^2(\Tcal;\lambda)
:=
\left\{
f:\Tcal\to\mathbb R:
\int_{\Tcal} f^2(t)\,d\lambda(t)<\infty
\right\},$
with inner product $\langle f,g\rangle_\lambda
:=
\int_{\Tcal} f(t)g(t)\,d\lambda(t)$
and norm $\|f\|_\lambda
:=
\left(\int_{\Tcal} f^2(t)\,d\lambda(t)\right)^{1/2}.$ The following examples illustrate how different choices of \(\lambda\) define downstream inferential targets.

\begin{example}[Barycenter quantile function and its linear functionals]
\label{ex:uniform_reference_cf}
Suppose \(\lambda\) is the uniform distribution on \([0,1]\). Then
$\theta^\lambda(u)
=
\E\{Y^{-1}(u)\}
=
\mu_\oplus^{-1}(u), u\in[0,1]$,which is the quantile function of the population Wasserstein
barycenter. For any
weight function \(\ell\in L^2[0,1]\), define $\theta_\ell^\lambda
:=
\langle \ell,\theta^\lambda\rangle_\lambda.$ A useful class of such functionals is obtained by considering a quantile region \(A=[p,q]\subset[0,1]\). Taking
\(\ell=\ell_A^\lambda\) gives $\theta_{\ell_A^\lambda}^\lambda
=
(q-p)^{-1}
\int_p^q \mu_\oplus^{-1}(u)\,du$ is the interquantile mean of the barycenter distribution over \(A\).
Special cases include the barycenter mean when \(A=[0,1]\), trimmed means when \(A=[\tau,1-\tau]\), and tail means when \(A=[\tau,1]\) or
\(A=[0,\tau]\), for some $\tau \in (0,1)$.
\end{example} 
\begin{example}[Displacement, Wasserstein distance, and transport map]
\label{ex:control_reference}
Let \(\mu_0\) denote a control distribution. For a continuous reference distribution \(\lambda\), Define the displacement function
\(D_{\mu_0}^\lambda:=\theta^\lambda-\mu_0^{-1}\circ\lambda
=\mu_\oplus^{-1}\circ\lambda-\mu_0^{-1}\circ\lambda\),
and, for any \(\ell\in L^2(\Tcal;\lambda)\), the displacement functional
\(\vartheta_\ell^\lambda:=\langle \ell,D_{\mu_0}^\lambda\rangle_\lambda\).
For a quantile region \(A=[p,q]\subset[0,1]\),
\(\vartheta_{\ell_A^\lambda}^\lambda
=(q-p)^{-1}\int_p^q
\{\mu_\oplus^{-1}(u)-\mu_0^{-1}(u)\}\,du\),
which includes mean displacement when \(A=[0,1]\) and tail displacement when \(A=[\tau,1]\) or \(A=[0,\tau]\). Moreover, $W_2(\mu_\oplus,\mu_0)=\|D_{\mu_0}^\lambda\|_\lambda.$ If \(\mu_0\) is continuous, it may itself be chosen as the bridge reference. In this case, $\theta^{\mu_0}=\mu_\oplus^{-1}\circ\mu_0$ is the optimal transport map from \(\mu_0\) to \(\mu_\oplus\) and $D_{\mu_0}^{\mu_0}=\theta^{\mu_0}-\mathrm{id}$ is the corresponding displacement map.
\end{example}

\begin{remark}[Degenerate scalar case]
\label{remark:degenerate_target_cf}
If the outcome distribution is degenerate, namely \(Y=\delta_V\) for a random variable \(V\), then \(Y^{-1}(u)=V\) for all \(u\in(0,1)\), and hence \(\theta^\lambda(\cdot)=\E[V]\). Therefore, the function-valued target \(\theta^\lambda\) reduces to the scalar mean.
\end{remark}

Examples \ref{ex:uniform_reference_cf} and  \ref{ex:control_reference} illustrate that the bridge parameter itself is not necessarily the final inferential target. Rather, it serves as a function-valued intermediate object from which a wide variety of scientifically meaningful quantities can be obtained through suitable functionals. This universality motivates our strategy of first constructing an efficient estimator for the bridge parameter and then deriving inference for downstream targets through the corresponding functionals.

\subsection{The generation powered inference for distributional bridge parameter under known generative surrogate}\label{subsec:basic-gpi}

We first motivate the construction under an idealized setting where the generative surrogate distribution \(G\) is directly available. Define the transformed outcome and transformed synthetic surrogate as $Z^\lambda(\cdot)
:=
Y^{-1}\circ\lambda(\cdot)$, $g^\lambda(X)(\cdot)
:=
G(X)^{-1}\circ\lambda(\cdot).$
By construction,
\begin{equation}\label{eq:theta-decomp}
\theta^\lambda
=
\E\!\left[\Gamma g^\lambda(X)\right]
+
\E\!\left[Z^\lambda-\Gamma g^\lambda(X)\right],
\end{equation}
where $\Gamma:L^2(\Tcal;\lambda)\to L^2(\Tcal;\lambda)$ is a bounded linear
operator controlling how information is borrowed from the generative surrogate \(g^\lambda(X)\) across \(\Tcal\). We define $R_\Gamma^\lambda
:=
\E\!\left[Z^\lambda-\Gamma g^\lambda(X)\right]$
as a distribution-level rectifier. The decomposition \eqref{eq:theta-decomp} into generative and rectification components is the distribution-valued analogue of the augmented representation in \eqref{eq:aug_eq_scalar}. We allow $\Gamma$ to be tuned for improved efficiency across downstream inferential tasks, which will be discussed in Section~\ref{subsec:choice-Gamma}.

\begin{remark}[Connection to PPI for finite-dimensional parameters]
The choice $\Gamma=I$ yields a distributional analogue of classical PPI \citep{Angelopoulos2023PPI}. More generally, the choices $\Gamma=\gamma I$ with $\gamma\in\mathbb{R}$, a multiplication operator, and a general linear operator provide distributional analogues of the scaled-identity, diagonal, and unrestricted matrix adjustments developed for finite-dimensional parameters in \citet{Angelopoulos2023PPIpp}, \citet{miao2025assumption}, and \citet{xu2025unified}, respectively.
\end{remark}

We provide details of constructing the GPI estimator based on the data observed and the surrogate generator $G$. For each \(i\in\mathcal O\), let \(\widehat Y_i\) denote an estimator of \(Y_i\) based on samples drawn from that distribution. Similarly, for any input \(x\), let \(\widehat G(x)\) denote an estimator of \(G(x)\) based on samples generated from it, and define the transformed surrogate $\hat g^\lambda(x):=\widehat G(x)^{-1}\circ\lambda.$ Then the basic GPI estimator is defined by
\begin{equation}
\label{eq:basiC_{gg}pi}
\tilde\theta^{\lambda,\Gamma}
:=
\frac1{N}\sum_{j\in\mathcal U}\Gamma \hat g^\lambda(X_j)
+
\frac1n\sum_{i\in\mathcal O}
\left\{
\widehat Y_i^{-1}\circ\lambda-\Gamma \hat g^\lambda(X_i)
\right\},
\end{equation}
where $
n^{-1}\sum_{i\in\mathcal O}
\{
\widehat Y_i^{-1}\circ\lambda-\Gamma \hat g^\lambda(X_i)
\}:=\hat R_{n,\Gamma}^\lambda$
can be viewed as the empirical rectifier.

\subsection{Cross-fitted generation powered inference under unknown generative surrogate}\label{subsec:cf-gpi}
In some applications, the generative surrogate \(G\) is not directly available and must be obtained by training a generative model or fine-tuning a pretrained model using the labeled data. In this setting, we propose a cross-fitted version of GPI. Let $\mathcal{O}_1,\ldots,\mathcal{O}_K$ be a random partition of the labeled sample into $K$ folds, and let $\mathcal{O}_{-k}:=\bigcup_{m\neq k}\mathcal{O}_m
$ denote the training sample for fold $k$.

For each $k=1,\ldots,K$, using only the training sample $\mathcal{O}_{-k}$, we train a fold-specific generative model. This induces a conditional distribution map $x \mapsto G_k(x),$
where $G_k(x)$ denotes the surrogate distribution generated under condition $x$ by the model trained on $\mathcal{O}_{-k}$. In general, $G_k(x)$ does not admit a closed-form quantile function. Therefore, for any specific input $x$, we generate Monte Carlo samples from $G_k(x)$ and let $\hat G_k(x)$ denote the corresponding empirical approximation. This induces the transformed surrogate $\hat g_k^{\lambda}(x):=\hat G_k(x)^{-1}\circ \lambda.$

Using the evaluation sample $\mathcal{O}_k$, we define the fold-specific estimator, for any bounded linear operator $\Gamma:L^2(\Tcal;\lambda)\to L^2(\Tcal;\lambda)$, by
$$\hat\theta_k^{\lambda,\Gamma}
=
N^{-1}\sum_{j\in\mathcal U}\Gamma \hat g_k^{\lambda}(X_j)
+
n_k^{-1}\sum_{i\in\mathcal O_k}
\left\{
\widehat Y_i^{-1}\circ\lambda-\Gamma \hat g_k^{\lambda}(X_i)
\right\},$$
where $n_k=|\mathcal{O}_k|$. Finally, we combine the fold-specific estimators via
\begin{equation}
\label{eq:cf_gpi}
\hat\theta^{\lambda,\Gamma}=
\sum_{k=1}^K \frac{n_k}{n}\,
\hat\theta_k^{\lambda,\Gamma}.
\end{equation}
Here \(\hat\theta^{\lambda,\Gamma}\) admits the decomposition $\hat\theta^{\lambda,\Gamma}=
\hat\theta^{\lambda,0}
+
\Gamma \widehat\Delta^{\lambda}$,
where $\hat\theta^{\lambda,0}$ is the estimator based on the labeled data only, and
$\widehat\Delta^{\lambda}
:=
\sum_{k=1}^K {n_k}/{n}
\{
{N}^{-1}\sum_{j\in\mathcal{U}} \hat g_k^{\lambda}(X_j)
-
{n_k}^{-1}\sum_{i\in\mathcal{O}_k}\hat g_k^{\lambda}(X_i)
\}$
is an empirical surrogate contrast whose population counterpart equals zero. Hence, the GPI estimator can be also viewed as the initial estimator plus a mean-zero correction term, in the spirit of \citet{YangDing2020}.

\section{Asymptotic efficiency of the GPI estimator}\label{sec:asymptotic}
In this section, we establish the asymptotic properties of the cross-fitted GPI estimator proposed in Section \ref{subsec:cf-gpi}. The corresponding results for the basic GPI estimator in Section \ref{subsec:basic-gpi} follow as
special cases. We introduce additional notation.

%For any bivariate function \(K:\Tcal\times\Tcal\to\mathbb R\), define $\|K\|_{\lambda}:=\{\int_{\Tcal}\int_{\Tcal}K^2(s,t)\,d\lambda(s)d\lambda(t)\}^{1/2}$. 
For a bounded linear operator
\(\Gamma:L^2(\Tcal;\lambda)\to L^2(\Tcal;\lambda)\), let \(\Gamma^\dagger\) denote its
adjoint, defined by
$\langle \Gamma f,g\rangle_\lambda
=
\langle f,\Gamma^\dagger g\rangle_\lambda$, and define $\|\Gamma\|_{\mathrm{op}}=\sup_{\|h\|_{\lambda}=1}\|\Gamma h\|_{\lambda}$. For a square-integrable covariance or cross-covariance kernel \(C\),
let \(\mathcal C\) denote the induced integral operator 
$(\mathcal C h)(s)
=
\int_{\Tcal}
C(s,t)h(t)\,d\lambda(t), 
h\in L^2(\Tcal;\lambda).$ For symmetric kernels \(C_1,C_2:\Tcal\times\Tcal\to\mathbb R\), write $C_1\preceq C_2 $ if \(C_2-C_1\) is positive semidefinite, that is, $\iint_{\Tcal\times\Tcal}
h(s)\{C_2(s,t)-C_1(s,t)\}h(t)
\,d\lambda(s)d\lambda(t)
\ge 0$
for all \(h\in L^2(\Tcal;\lambda)\). Denote weak convergence by \(\rightsquigarrow\) and convergence in probability by \(\overset{p}{\longrightarrow}\).

%For any \(L^2(\Tcal;\lambda)\)-valued function \(h(x)\), define its integrated squared norm by
%\[
%\mid\!\mid\!\mid h \mid\!\mid\!\mid_\lambda^2
%:=
%\int \|h(x)\|_\lambda^2\, dF_X(x),
%\]
%where \(F_X\) denotes the probability measure induced by \(X\).

%We make the following assumptions.

\begin{assumption}[Distribution estimation]
\label{ass:dist}
The following conditions hold.

\begin{enumerate}
\item[(a)] The estimators $\widehat Y_1,\ldots,\widehat Y_n$ are independent. For some deterministic sequences $\alpha_n=o(1)$ and $\nu_n=o(1)$, $$\sup_{1\le i\le n}\sup_{\upsilon\in\W_2(\Ical)}
\E\!\{
W_2^2(\widehat Y_i,Y_i)\mid Y_i=\upsilon
\}
=
O(\alpha_n^2),$$ and $$
\sup_{1\le i\le n}\sup_{\upsilon\in\W_2(\Ical)}
\Var\!\{
W_2^2(\widehat Y_i,Y_i)\mid Y_i=\upsilon
\}
=
O(\nu_n^4).
$$

\item[(b)] For each fold \(k\), the estimators $\{\widehat{G}_k(X_i)\}_{i\in\mathcal{O}_k}$ are independent conditional on $\mathcal{O}_{-k}$ and
$$\sup_{i\in\mathcal{O}_k}\sup_{\upsilon\in\W_2(\Ical)}
\E\!\{
W_2^2\!\bigl(\hat G_k(X_i),G_k(X_i)\bigr)
\;|\; \mathcal O_{-k}, G_k(X_i)=\upsilon
\}
=
O_p\!\bigl(\alpha_n^2\bigr),$$ and $$
\sup_{i\in\mathcal{O}_k}\sup_{\upsilon\in\W_2(\Ical)}
\Var\!\{
W_2^2\!\bigl(\hat G_k(X_i), G_k(X_i)\bigr)
\;|\; \mathcal O_{-k}, G_k(X_i)=\upsilon
\}
=
O_p\!\bigl(\nu_n^4\bigr).$$
The same bounds hold for unlabeled points \(X_j\), \(j\in\mathcal{U}\), with \(n\) replaced by \(N\).
\end{enumerate}
\end{assumption}

Assumption~\ref{ass:dist} specifies convergence rates for estimating the observed distributions (a) and the generated distributions (b), as in \citet{lin2022causal}. When each distribution is estimated from \(m_i\asymp n^\zeta\) observations, the empirical distribution satisfies \(\alpha_n^2=\nu_n^4=n^{-\zeta/2}\) under suitable moment assumptions \citep{fournier2015rate}. With additional regularity \citep{petersen2016functional}, a nonparametric distribution estimator can achieve the faster rates \(\alpha_n^2=n^{-2\zeta/3}\) and \(\nu_n^4=n^{-4\zeta/3}\).

%\begin{assumption}[Reference distribution]
%\label{ass:lam}
%The global reference estimator $\hat\lambda$ and the fold-specific estimators $\hat\lambda$ satisfy
%\[
%W_2^2(\hat\lambda,\lambda)=O_p(n^{-1}+\alpha_n^%2+\nu_n^2).
%\]
%If $\lambda$ is fixed, this assumption holds %trivially.
%\end{assumption}

Further, for each fold \(k\), let \(\tilde G_k(x)\) denote the population target of the fold-specific generative fitting procedure when the true distributions \(Y_i\) are observed, and let \(G^*(x)\) denote its working-model target as the training sample size tends to infinity. Define the oracle transformed surrogate outcome $
g^{\lambda,*}(x):=G^*(x)^{-1}\circ\lambda$.

\begin{assumption}[Generative model]
\label{ass:g}
Conditional on the training fold \(\mathcal O_{-k}\), the fold-specific population generative targets satisfy (a) $\E_X\!\left[
    W_2^2\!\bigl(G_k(X),\tilde G_k(X)\bigr)
    \,\middle|\, \mathcal O_{-k}
    \right]
    =
    O_p(\alpha_n^2+\nu_n^2),$ for $k=1,\ldots,K;$
    (b) $\E_X\!\left[
    W_2^2\!\bigl(\tilde G_k(X),G^*(X)\bigr)
    \,\middle|\, \mathcal O_{-k}
    \right]
    =
    O_p(r_g^2),$ for $k=1,\ldots,K;$
    and (c) $\E\|g^{\lambda,*}(X)\|_{\lambda}^2<\infty.$ Here \(\E_X\) denotes expectation with respect to the marginal distribution of the covariate \(X\).
\end{assumption}

Assumption~\ref{ass:g} decomposes the generative-model error into two sources: \ref{ass:g}(a) captures the effect of estimating the labeled distributions from finite samples, while \ref{ass:g}(b) captures the discrepancy between the generative target trained on finitely many labeled distributions and its population limit $G^*$. \ref{ass:g}(c) imposes a finite second-moment condition on the bridge representation of the generative surrogate.

\begin{assumption}[Fold sizes]
\label{ass:fold}
There exist constants $0<c_1\le c_2<1$ such that $c_1\le {n_k}/{n}\le c_2$
for all $k=1,\ldots,K$.
\end{assumption}

\subsection{Asymptotic efficiency and optimal choice of $\Gamma$}\label{subsec:choice-Gamma}
Define $\phi_\Gamma^\lambda(X,Y)
:=
Z^\lambda-\Gamma g^{\lambda,*}(X)$ and $
\psi_\Gamma^\lambda(X)
:=
\Gamma g^{\lambda,*}(X).$ Let
$\mathbb G_n f
:= n^{-1/2}
\sum_{i\in\mathcal O}
\left[
f(X_i,Y_i)\right.$ $\left.-\E\{f(X,Y)\}
\right]$, $
\widetilde{\mathbb G}_N h
:=N^{-1/2}
\sum_{j\in\mathcal U}
\left[
h(X_j)-\E\{h(X)\}
\right],$
denote the labeled and unlabeled empirical processes, respectively. The following theorem establishes the asymptotic properties of the cross-fitted GPI estimator.

\begin{theorem}[Fixed-$\Gamma$ asymptotics]
\label{thm:fixed}
Let \(\Gamma:L^2(\Tcal;\lambda)\to L^2(\Tcal;\lambda)\) be a fixed bounded linear operator. Suppose Assumptions \ref{ass:dist}-\ref{ass:fold} hold and $n/N\to\rho\in[0,\infty)$. Then $\|
\hat\theta^{\lambda,\Gamma}
-
\theta^\lambda
\|_{\lambda}
=
O_p\!(
\alpha_n+\nu_n+n^{-1/2}r_g+n^{-1/2}
).$
In particular, if \(\alpha_n=o(1)\), \(\nu_n=o(1)\), and \(r_g=O_p(1)\), then
$\|
\hat\theta^{\lambda,\Gamma}
-
\theta^\lambda
\|_{\lambda}
=o_p(1).$ If, in addition, \(\alpha_n+\nu_n=o(n^{-1/2})\) and \(r_g=o_p(1)\), then
\begin{equation}
\label{eq:lin-fixed}
\sqrt n\left(
\hat\theta^{\lambda,\Gamma}
-
\theta^\lambda
\right)
=
\mathbb G_n\,\phi_\Gamma^\lambda
+
\sqrt{\frac nN}\,\widetilde{\mathbb G}_N\,\psi_\Gamma^\lambda
+o_p(1)
\quad\text{in }L^2(\Tcal;\lambda).
\end{equation}
Consequently, 
$\sqrt n(
\hat\theta^{\lambda,\Gamma}
-
\theta^\lambda)
\rightsquigarrow
\mathbb G_\Gamma^\lambda
\quad\text{in }L^2(\Tcal;\lambda),$
where \(\mathbb G_\Gamma^\lambda\) is a centered Gaussian random element with covariance kernel
\begin{align}
\label{eq:cov-fixed}
C_\Gamma^\lambda(s,t)
&=
\Cov\!\bigl(
Z^\lambda(s)-(\Gamma g^{\lambda,*}(X))(s),\,
Z^\lambda(t)-(\Gamma g^{\lambda,*}(X))(t)
\bigr)
\nonumber\\
&\qquad
+
\rho\,
\Cov\!\bigl(
(\Gamma g^{\lambda,*}(X))(s),\,
(\Gamma g^{\lambda,*}(X))(t)
\bigr),
\qquad s,t\in\Tcal.
\end{align}
\end{theorem}

Theorem~\ref{thm:fixed} shows that \(\hat\theta^{\lambda,\Gamma}\) converges to a Gaussian process \(\mathbb G_\Gamma^\lambda\) with covariance kernel \(C_\Gamma^\lambda\), so the choice of \(\Gamma\) determines asymptotic efficiency. For two operators \(\Gamma_1\) and \(\Gamma_2\), we say
that \(\Gamma_1\) is asymptotically no less efficient than \(\Gamma_2\) if
$C_{\Gamma_1}^\lambda \preceq C_{\Gamma_2}^\lambda.$ Equivalently, every linear functional
\(\langle a,\hat\theta^{\lambda,\Gamma}\rangle_\lambda\) has
an asymptotic variance under \(\Gamma_1\) no larger than that under
\(\Gamma_2\). Define $C_{gg}^\lambda(s,t)
=
\Cov\!\bigl(g^{\lambda,*}(X)(s),g^{\lambda,*}(X)(t)\bigr)$,  $C_{Zg}^\lambda(s,t)
=
\Cov\!\bigl(Z^\lambda(s),g^{\lambda,*}(X)(t)\bigr), s,t\in\Tcal$ and let \(\mathcal C_{gg}^\lambda\),
\(\mathcal C_{Zg}^\lambda\), and \(\mathcal C_\Gamma^\lambda\) denote the
operators induced by \(C_{gg}^\lambda\), \(C_{Zg}^\lambda\), and
\(C_\Gamma^\lambda\), respectively. For the labeled-only estimator, write
\(C_0^\lambda:=C_{\Gamma=0}^\lambda\) and
\(\mathcal C_0^\lambda:=\mathcal C_{\Gamma=0}^\lambda\). We have the following theorem.

\begin{theorem}[Optimal choice of \(\Gamma\) and efficiency gain]
\label{thm:Gamma-star}
Under the conditions of Theorem \ref{thm:fixed}, let
\(\Gamma_*:L^2(\Tcal;\lambda)\to L^2(\Tcal;\lambda)\) be a bounded linear operator satisfying
\begin{equation}
\label{eq:Gamma-star-normal-kernel}
(1+\rho)\Gamma_*\mathcal C_{gg}^\lambda
=
\mathcal C_{Zg}^\lambda.
\end{equation}
Then \(\Gamma_*\) is covariance-operator optimal over the class of bounded
linear operators on \(L^2(\Tcal;\lambda)\), in the sense that $C_{\Gamma_*}^\lambda
\preceq
C_\Gamma^\lambda$
for every such operator \(\Gamma\). More precisely, letting
\(\mathcal C_\Gamma^\lambda\) denote the covariance operator induced by
\(C_\Gamma^\lambda\),
\begin{equation}
\label{eq:operator-efficiency-gain}
\mathcal C_\Gamma^\lambda
-
\mathcal C_{\Gamma_*}^\lambda
=
(1+\rho)
(\Gamma-\Gamma_*)
\mathcal C_{gg}^\lambda
(\Gamma-\Gamma_*)^\dagger.
\end{equation}
In particular, compared with the labeled-only estimator corresponding
to \(\Gamma=0\), \(C_{\Gamma_*}^\lambda\preceq C_0^\lambda\), and the covariance-operator efficiency gain is $\mathcal C_0^\lambda
-
\mathcal C_{\Gamma_*}^\lambda
=
(1+\rho)
\Gamma_*\mathcal C_{gg}^\lambda\Gamma_*^\dagger.$
\end{theorem}

Theorem \ref{thm:Gamma-star} interprets the optimal borrowing operator
\(\Gamma_*\) as a shrunk functional regression operator of the transformed outcome \(Z^\lambda\) on the transformed surrogate \(g^{\lambda,*}(X)\), with the shrinkage accounting for variability from the unlabeled surrogate average. The following corollary gives a particularly simple form of the oracle operator \(\Gamma_*\)
when the generative surrogate is a a conditional projection of \(Z^\lambda\).
\begin{corollary}\label{cor:simplified-case}
Suppose the conditions of Theorem \ref{thm:Gamma-star} hold. If the
generative surrogate satisfies $g^{\lambda,*}(X)
=
\E[Z^\lambda\mid h(X)]$
for some feature map \(h:\mathbb{R}^d \rightarrow\mathbb{R}^r\) with $r\in \mathbb{N}^{+}$, then $C_{Zg}^\lambda(s,t)=C_{gg}^\lambda(s,t),$ for \(\lambda\otimes\lambda\)-almost every
\((s,t)\in\Tcal^2\).
Consequently, one valid oracle operator is $\Gamma_*=(1+\rho)^{-1}I.$ Moreover,
$\mathcal C_0^\lambda-\mathcal C_{\Gamma_*}^\lambda
=(1+\rho)^{-1}\mathcal C_{gg}^\lambda.$
\end{corollary}
More generally, Theorem~\ref{thm:Gamma-star} shows that the efficiency gain is
governed by the variation in \(Z^\lambda\) that is linearly explained by the generative surrogate. Thus, the efficiency gain increases with surrogate
informativeness and with the relative size of the unlabeled sample. Two extreme cases are worth noting.

First, suppose the transformed outcome is perfectly linearly recoverable from the generative surrogate. That is, there exists a bounded linear operator \(\Gamma_0:L^2(\Tcal;\lambda)\to L^2(\Tcal;\lambda)\) such that $ \Gamma_0\{g^{\lambda,*}(X)-\E g^{\lambda,*}(X)\} = Z^\lambda-\E Z^\lambda $ a.s., then the oracle borrowing operator can be chosen as $\Gamma_*={(1+\rho)}^{-1}\Gamma_0$, and $C_{\Gamma_*}^\lambda(s,t) ={\rho}/{(1+\rho)}C_0^\lambda(s,t), s,t\in\Tcal,$ yielding the maximal covariance-kernel efficiency gain. 

Second, if the generative surrogate is linearly uninformative in the sense that $C_{Zg}^\lambda(s,t)=0, s,t\in\Tcal,$ then \(\Gamma_*=0\) is an oracle solution of \eqref{eq:Gamma-star-normal-kernel}, and $C_{\Gamma_*}^\lambda=C_0^\lambda.$ Thus, a non-informative surrogate produces no asymptotic efficiency loss compared to the labeled-only estimator, reflecting the robustness of the GPI estimator.

%As illustrated in Examples \ref{ex:uniform_reference_cf}--\ref{ex:control_reference}, many inferential quantities derived from the distributional target can be written as linear functionals of \(\theta^\lambda\). Since covariance-kernel optimality implies optimality for every fixed linear functional, the following corollary translates this efficiency gain into an effective sample size for such quantities.

\subsection{Estimation of the optimal operator}\label{subsec:Gamma-estimation}
In the simplified case of Corollary \ref{cor:simplified-case}, the oracle
operator can be estimated by
\((1+\hat\rho)^{-1}I\), where \(\hat\rho=n/N\). For general black-box
generative models, this projection structure may not hold. For feasible estimation, we assume that \(\Gamma_*\) admits an integral representation
$(\Gamma_*f)(t)
=
\int_{\Tcal}
\Gamma_*(t,s)f(s)\,d\lambda(s), 
f\in L^2(\Tcal;\lambda),$
for a bivariate function $\Gamma_*(t,s):\Tcal\times\Tcal\to\mathbb R$. We provide an
estimation procedure based on functional principal component analysis (FPCA) for \(\Gamma_*\) characterized by
\eqref{eq:Gamma-star-normal-kernel}.

For each labeled observation \(i\in\mathcal O_k\), define $\hat Z_i^\lambda(t):=\widehat Y_i^{-1}\circ\lambda(t)$, $\hat g_i^\lambda(t):=\hat g_k^\lambda(X_i)(t), t\in\Tcal.$ For \(j\in\mathcal U\), define $\hat g_j^\lambda(t)
:=
\sum_{k=1}^K \omega_k \hat g_k^\lambda(X_j)(t)$, with $
\omega_k:={n_k}/{n}.$
Let \(\mathcal D=\mathcal O\cup\mathcal U\) and write
\(\hat g_i^\lambda\) for all \(i\in\mathcal D\). Define \(\bar Z^\lambda(t):=n^{-1}\sum_{i\in\mathcal O}\hat Z_i^\lambda(t)\) and
\(\bar g_{\mathcal D}^\lambda(t):=(n+N)^{-1}\sum_{i\in\mathcal D}\hat g_i^\lambda(t)\). We estimate the covariance kernels by
\(\hat C_{ZZ}^\lambda(s,t):=
n^{-1}\sum_{i\in\mathcal O}
\{\hat Z_i^\lambda(s)-\bar Z^\lambda(s)\}
\{\hat Z_i^\lambda(t)-\bar Z^\lambda(t)\}\)
and 
\(\hat C_{gg}^\lambda(s,t):=
(n+N)^{-1}\sum_{i\in\mathcal D}
\{\hat g_i^\lambda(s)-\bar g_{\mathcal D}^\lambda(s)\}
\{\hat g_i^\lambda(t)-\bar g_{\mathcal D}^\lambda(t)\}\). We eigendecompose \(\hat C_{ZZ}^\lambda\) and \(\hat C_{gg}^\lambda\) as
$\hat C_{ZZ}^\lambda(s,t)=\sum_{\ell\ge 1}\hat a_\ell
\hat u_\ell(s)\hat u_\ell(t)$ and $
\hat C_{gg}^\lambda(s,t)=\sum_{m\ge 1}\hat d_m
\hat v_m(s)\hat v_m(t),$
where \(\hat a_1\ge \hat a_2\ge\cdots\ge 0\) and
\(\hat d_1\ge \hat d_2\ge\cdots\ge 0\). For selected truncation levels
\(L\) and \(M\), define the FPC scores
$\hat\eta_{i\ell}
:=
\int_{\Tcal}
\{\hat Z_i^\lambda(t)-\bar Z^\lambda(t)\}\hat u_\ell(t)\,d\lambda(t),
i\in\mathcal O, \ell=1,\ldots,L,$
and
$\hat\xi_{im}
:=
\int_{\Tcal}
\{\hat g_i^\lambda(t)-\bar g_{\mathcal D}^\lambda(t)\}\hat v_m(t)\,d\lambda(t),
i\in\mathcal D, m=1,\ldots,M.
$
The estimated bivariate coefficient function of the optimal borrowing operator is given by
\begin{equation}\label{eq:Gammahat}
\hat\Gamma_*(t,s)
=
\frac{1}{1+\hat\rho}
\sum_{\ell=1}^{L}\sum_{m=1}^{M}
\hat b_{\ell m}\hat u_\ell(t)\hat v_m(s),
\qquad
\hat\rho=\frac nN,
\end{equation}
where $\hat b_{\ell m}
= {\hat{\varsigma}_{\ell m}}/{(\hat d_m+\kappa)},$ and $ \hat{\varsigma}_{\ell m}=n^{-1}\sum_{i\in \mathcal{O}}\hat\eta_{i\ell}\hat{\xi}_{im},$ where \(\kappa\ge 0\) is a ridge regularization parameter. Details on selecting \(L\), \(M\), and \(\kappa\) are provided in Supplementary Section~\ref{sec:implement-detail}. The estimated coefficient function induces the linear operator \(\hat\Gamma_*:L^2(\Tcal;\lambda)\to
L^2(\Tcal;\lambda)\) through
$(\hat\Gamma_* f)(t)
=
\int_{\Tcal}
\hat\Gamma_*(t,s)f(s)\,d\lambda(s), 
f\in L^2(\Tcal;\lambda).$ The resulting feasible, optimally tuned estimator is given by $
\hat\theta^{\lambda,\hat\Gamma_*}$. 

\begin{assumption}[Operator estimation]
\label{ass:Gamma-consistency} $\hat\Gamma_*$ satisfies
$\|\hat\Gamma_*-\Gamma_*\|_{\mathrm{op}}=o_p(1)$.
\end{assumption}

Assumption~\ref{ass:Gamma-consistency} can be verified using existing consistency results for coefficient-kernel estimation in
functional linear regression \citep{YaoMullerWang2005}, under suitable
smoothness and eigenvalue-decay, and truncation-rate conditions. The following proposition characterizes the asymptotic
behavior of the feasible estimator.

\begin{proposition}[Asymptotic equivalence of the feasible estimator]
\label{prop:feasible-oracle}
Suppose Assumptions~\ref{ass:dist}-\ref{ass:Gamma-consistency} hold,
$n/N\to\rho\in[0,\infty)$,
$\alpha_n+\nu_n=o(n^{-1/2})$, and $r_g=o_p(1)$. Then
$\sqrt n\,
\bigl\|
\hat\theta^{\lambda,\hat{\Gamma}_*}
-
\hat\theta^{\lambda,\Gamma_*}
\bigr\|_\lambda
=
o_p(1).$
Consequently, $\sqrt n
\bigl(
\hat\theta^{\lambda,\hat{\Gamma}_*}-\theta^\lambda
\bigr)
\rightsquigarrow
\mathbb G_{\Gamma_*}^\lambda, \text{ in } L^2(\Tcal;\lambda).$
\end{proposition}

\section{Inference for the bridge parameter and its functionals}\label{sec:inference}
\subsection{Inference on the bridge parameter}\label{subsec:bridge-inference}
With the estimator \(\hat\Gamma_*\) in \eqref{eq:Gammahat}, we can estimate the covariance kernel associated with the oracle-tuned estimator based on the decomposition in \eqref{eq:cov-fixed}. To mitigate in-sample bias, we use cross-fitted borrowed surrogate components in the labeled residuals. Specifically, let
\(\hat h_{i,\mathrm{cf}}^\lambda\) denote the cross-fitted estimate of
\(h_*^\lambda(X_i):=\Gamma_*g^{\lambda,*}(X_i)\) for \(i\in\mathcal O\), obtained by
estimating \(\Gamma_*\) without using the fold containing observation
\(i\). Details of this cross-fitting construction are provided in Supplementary Section~\ref{sec:implement-detail}.
For the second covariance term, we use the full-sample estimator
\(\hat\Gamma_*\) and define
\(\hat h_i^\lambda:=\hat\Gamma_*\hat g_i^\lambda\) for
\(i\in\mathcal D\). We then estimate $C_{\Gamma_*}^\lambda$ by
\begin{equation*}
\label{eq:Cstarhat}
\hat C_*^\lambda(s,t)
:=
\widehat{\Cov}_{\mathcal O}
\{\hat Z_i^\lambda(s)-\hat h_{i,\mathrm{cf}}^\lambda(s),\hat Z_i^\lambda(t)-\hat h_{i,\mathrm{cf}}^\lambda(t)\}
+
\hat\rho\,
\widehat{\Cov}_{\mathcal D}
\{\hat h_i^\lambda(s),\hat h_i^\lambda(t)\}, s,t\in\Tcal,
\end{equation*}
where \(\widehat{\Cov}_{\mathcal O}\) and
\(\widehat{\Cov}_{\mathcal D}\) denote the empirical covariance kernels
computed from the labeled and pooled samples, respectively.

By construction,
\(\hat C_*^\lambda\) is positive semidefinite and therefore determines,
conditionally on the observed data, a centered Gaussian process
\(\hat{\mathbb G}_*^\lambda\) on \(\Tcal\) with covariance kernel
\(\hat C_*^\lambda\). We generate
\(B\) independent sample paths
$\hat{\mathbb G}_{1,*}^\lambda,\ldots,
\hat{\mathbb G}_{B,*}^\lambda$
from this conditional Gaussian process and compute $S_b
:=
\sup_{t\in\Tcal}
|\hat{\mathbb G}_{b,*}^\lambda(t)|, b=1,\ldots,B.$
Let \(\hat q_{1-\alpha,*}\) be the empirical
\((1-\alpha)\)-quantile of \(S_1,\ldots,S_B\). A Gaussian
approximation-based simultaneous confidence band (SCB) for
\(\theta^\lambda\) is then given by
\begin{equation}
\label{eq:scb}
\left[
\hat\theta^{\lambda,\hat{\Gamma}_*}(t)-n^{-1/2}\hat q_{1-\alpha,*},
\;
\hat\theta^{\lambda,\hat{\Gamma}_*}(t)+n^{-1/2}\hat q_{1-\alpha,*}
\right],
\qquad t\in\Tcal.
\end{equation}
Its width provides a curve-level measure of uncertainty. The following result establishes the consistency of the resulting Monte Carlo critical value.

\begin{theorem}
\label{thm:conditional-gaussian-approximation}

Under Assumption~\ref{ass:conditional-gaussian-approximation} in Supplementary Section \ref{subsec:add-theory-ga}, if the distribution function of $\sup_{t\in\Tcal}|\mathbb G_{\Gamma_*}^\lambda(t)|$ is continuous and strictly increasing in a neighborhood of its \((1-\alpha)\)-quantile
\(q_{1-\alpha,*}\), and \(B\to\infty\), then $\hat q_{1-\alpha,*}\overset{p}{\longrightarrow}q_{1-\alpha,*}$.
\end{theorem}

\subsection{Inference on linear functionals}\label{subsec:linear-functional}
We next consider inference for fixed linear functionals of the bridge parameter. For \(\ell\in L^2(\Tcal;\lambda)\), define
\(\theta_\ell^\lambda
:=
\langle \ell,\theta^\lambda\rangle_\lambda\)
and 
\(\hat\theta_{\ell}^{\lambda,\Gamma}
:=
\langle \ell,\hat\theta^{\lambda,\Gamma}\rangle_\lambda\).
As shown in Examples~\ref{ex:uniform_reference_cf}--\ref{ex:control_reference},
this class covers several common barycenter summaries, with displacement functionals differing only by a known deterministic term. The oracle-tuned and labeled-only estimators are
\(\hat\theta_{\ell}^{\lambda,\Gamma_*}\) and
\(\hat\theta_{\ell}^{\lambda,0}\), respectively.

\begin{corollary}[Asymptotic efficiency for linear functionals]
\label{cor:linear-functional}
Suppose the conditions of Theorem~\ref{thm:fixed} hold. Then, for any fixed
\(\ell\in L^2(\Tcal;\lambda)\) and any fixed bounded linear operator $\Gamma$,$\sqrt n
(
\hat\theta_{\ell}^{\lambda,\Gamma}
-
\theta_\ell^\lambda
)
\rightsquigarrow
N\!(
0,
V_{\Gamma,\ell}^\lambda
),$
where $V_{\Gamma,\ell}^\lambda
:=
\Var\{\langle
\ell,
\mathbb G_{\Gamma}^\lambda
\rangle_\lambda
\}.$ If, in addition, the conditions of
Theorem~\ref{thm:Gamma-star} hold, then the optimal operator \(\Gamma_*\) defined in \eqref{eq:Gamma-star-normal-kernel} yields the oracle asymptotic variance
\(V_{\Gamma_*,\ell}^\lambda\), and is simultaneously optimal
for all fixed linear functionals. Specifically, for any fixed bounded linear operator
\(\Gamma\), 
we have $V_{\Gamma,\ell}^\lambda
-
V_{\Gamma_*,\ell}^\lambda
=
(1+\rho)
\Var\{
\langle
\ell,
(\Gamma-\Gamma_*)g^{\lambda,*}(X)
\rangle_\lambda
\}
\ge 0.$
In particular, let $V_{0,\ell}^\lambda
:=
\Var\{
\langle
\ell,
\mathbb G_{0}^\lambda
\rangle_\lambda
\}$
denote the asymptotic variance of the labeled-only estimator
\(\hat\theta_{\ell}^{\lambda,0}\). Then $V_{0,\ell}^\lambda
-
V_{\Gamma_*,\ell}^\lambda
=
(1+\rho)
\Var\{
\langle
\ell,
\Gamma_*g^{\lambda,*}(X)
\rangle_\lambda
\}
\ge 0.$
\end{corollary}
\begin{remark}\label{rm:eff-n}
    If \(V_{\Gamma_*,\ell}^\lambda>0\), define the effective sample size by $n_{\mathrm{eff}}^\lambda(\ell)
:=
n{V_{0,\ell}^\lambda}/{V_{\Gamma_*,\ell}^\lambda}.$
Then \(n_{\mathrm{eff}}^\lambda(\ell)\ge n\). In the linearly
uninformative case, \(n_{\mathrm{eff}}^\lambda(\ell)=n\). In the
perfectly linearly informative case, for any \(\ell\) with
\(V_{0,\ell}^\lambda>0\),$
n_{\mathrm{eff}}^\lambda(\ell)
=
n{(1+\rho)}/{\rho}$, for $\rho>0.$
\end{remark}
For feasible inference based on
\(\hat\theta_{\ell}^{\lambda,\hat{\Gamma}_*}\), we estimate the
asymptotic variance \(V_{\Gamma_*,\ell}^\lambda\) by
$\hat V_{\ell}^\lambda
:=
\int_{\Tcal}\int_{\Tcal}
\ell(s)\hat C_*^\lambda(s,t)\ell(t)\,
d\lambda(s)d\lambda(t),$ where \(\hat C_*^\lambda\) is defined in Section \ref{subsec:bridge-inference}.
Under regularity conditions in
Supplementary Section~\ref{subsec:add-theory-lf}, we show that, provided
\(V_{\Gamma_*,\ell}^\lambda>0\),
$\sqrt n
\{\hat V_\ell^\lambda\}^{-1/2}
(
\hat\theta_{\ell}^{\lambda,\hat{\Gamma}_*}
-
\theta_\ell^\lambda
)
\rightsquigarrow
N(0,1).$ Consequently, an asymptotically valid
\(100(1-\alpha)\%\) confidence interval (CI) for \(\theta_\ell^\lambda\) is
\begin{equation}\label{eq:lf-CI}
\left[
\hat\theta_{\ell}^{\lambda,\hat{\Gamma}_*}
-
z_{1-\alpha/2}n^{-1/2}
\{\hat V_{\ell}^\lambda\}^{1/2},
\;
\hat\theta_{\ell}^{\lambda,\hat{\Gamma}_*}
+
z_{1-\alpha/2}n^{-1/2}
\{\hat V_{\ell}^\lambda\}^{1/2}
\right].
\end{equation}

\subsection{Inference on Wasserstein distance}
\label{subsec:w2-test}
In this section, we consider inference on the Wasserstein distance based on the bridge parameter. Let \(\mu_0\in\mathcal W_2(\mathcal I)\) be a fixed control distribution for comparison. Define $\Psi_{\mu_0}^{\lambda}(f):=\left\|f-\mu_0^{-1}\circ\lambda\right\|_{\lambda}.$ By Proposition~\ref{prop:w2-representation}, $\Psi_{\mu_0}^{\lambda}(\theta^\lambda)=W_2(\mu_0,\mu_\oplus).$

\begin{theorem}[Inference on $W_2$ distance]
\label{thm:w2}
Suppose the conditions of Theorem~\ref{thm:fixed} hold, and let \(\Gamma\) be any fixed bounded linear operator. Then 

(i) $\Psi_{\mu_0}^\lambda\!\left(\hat\theta^{\lambda,\Gamma}\right)\overset{p}{\longrightarrow}W_2(\mu_0,\mu_\oplus)$; 

(ii) if $W_2(\mu_0,\mu_\oplus)=0$, then $\sqrt n\,\Psi_{\mu_0}^\lambda\!\left(\hat\theta^{\lambda,\Gamma}\right)\rightsquigarrow\|\mathbb G_\Gamma^\lambda\|_\lambda$; 

(iii) if $W_2(\mu_0,\mu_\oplus)>0$, then $\sqrt n\left\{\Psi_{\mu_0}^\lambda\!\left(\hat\theta^{\lambda,\Gamma}\right)-W_2(\mu_0,\mu_\oplus)\right\}\rightsquigarrow N(0,\sigma_\Gamma^2),$ where $\sigma_\Gamma^2=V_{\Gamma,D_{\mu_0}^\lambda}^{\lambda}/W_2^2(\mu_0,\mu_\oplus)$ and $D_{\mu_0}^\lambda$ is the quantile displacement defined in Example~\ref{ex:control_reference}. If, in addition, the conditions of Theorem~\ref{thm:Gamma-star} hold, then the optimal operator \(\Gamma_*\) defined in \eqref{eq:Gamma-star-normal-kernel} satisfies $\sigma_{\Gamma_*}^2\le \sigma_\Gamma^2$. In particular, $\sigma_{\Gamma_*}^2\le \sigma_0^2,$ where \(\sigma_0^2\) is the asymptotic variance of the labeled-only estimator. Moreover, $\sigma_0^2-\sigma_{\Gamma_*}^2=(1+\rho)\Var\!\left(\left\langle D_{\mu_0}^\lambda,\Gamma_*g^{\lambda,*}(X)\right\rangle_\lambda\right)/W_2^2(\mu_0,\mu_\oplus)\ge 0.$
\end{theorem}

Theorem~\ref{thm:w2} also yields a natural test for equality between the control
distribution $\mu_0$ and the barycenter distribution $\mu_\oplus$. Consider the hypothesis testing
problem $H_0:\ W_2(\mu_0,\mu_\oplus)=0
\text{ versus }
H_1:\ W_2(\mu_0,\mu_\oplus)>0.$
Motivated by Theorem~\ref{thm:w2}(ii), define the test statistic $T_{n,*}
:=
\sqrt n\,
\Psi_{\mu_0}^{\lambda}
(\hat\theta^{\lambda,\hat\Gamma_*}),$
where \(\hat\Gamma_*\) is constructed as proposed in
Section~\ref{subsec:Gamma-estimation}. Under \(H_0\),
Theorem~\ref{thm:w2}(ii) and Proposition~\ref{prop:feasible-oracle} give
$T_{n,*}
\rightsquigarrow
\left\|
\mathbb G_{\Gamma_*}^\lambda
\right\|_\lambda.$ As in Section~\ref{subsec:Gamma-estimation}, generate \(B\) independent sample paths
$\hat{\mathbb G}_{1,*}^{\lambda},
\ldots,
\hat{\mathbb G}_{B,*}^{\lambda}$ from the centered Gaussian process on \(\Tcal\) with covariance kernel \(\hat C_*^\lambda\) defined in Section~\ref{subsec:bridge-inference}. For each \(b=1,\ldots,B\), compute
$S_b
:=\|
\hat{\mathbb G}_{b,*}^{\lambda}\|_\lambda,$
and let \(\hat q_{1-\alpha,*}\) be the empirical \((1-\alpha)\)-quantile of
\(S_1,\ldots,S_B\). We reject \(H_0\) at significance level \(\alpha\) whenever
$T_{n,*}>\hat q_{1-\alpha,*}.$
The labeled-only test is obtained by setting \(\Gamma=0\), in which case no
generative rectification is applied. Under any fixed alternative satisfying $W_2(\lambda,\mu_\oplus)>0$, the corresponding test is consistent, and
hence its power converges to one. Theorem \ref{thm:w2}(iii) shows that the optimal GPI
estimator has no larger asymptotic variance than the labeled-only estimator, indicating a potential finite-sample power gain through more precise estimation of the Wasserstein distance.

\section{Simulation Studies}\label{sec:simu}
\subsection{Simulation setup and data generation}\label{subsec:simu-setup}

We generate distribution-valued outcomes through quantile functions. Let \(u\in(0,1)\) denote the quantile level. The outcome barycenter is specified by the bounded, right-skewed quantile curve, $\mu_{\oplus}^{-1}(u)=0.1+ 0.8(z_\alpha(u)-z_\alpha(0))/(z_\alpha(1)-z_\alpha(0))$, where $z_\alpha(u)
=
F_{\mathrm{SN}}^{-1}\{\varepsilon+(1-2\varepsilon)u;\alpha\},
$ and \(F_{\mathrm{SN}}(\cdot;\alpha)\) is the skew-normal distribution function. We set \(\alpha=4\) and \(\varepsilon=10^{-3}\). For each subject \(i\), we generate 
\(X_i\in[-1,1]^3\), with entries independently drawn from
\(2\mathrm{Beta}(0.5,0.5)-1\). The subject-specific outcome quantile function is
defined by $Y_i^{-1}(u)
=
\mu^{-1}_\oplus(u)
+
\sum_{k=1}^{3}\eta_{ik}\phi_k(u),$
where \(\phi_1(u)=1\), \(\phi_2(u)=\sqrt{12}(u-1/2)\), and \(\phi_3(u)=\sqrt5(6u^2-6u+1)\) are orthonormal functions in \(L^2[0,1]\), representing a location shift, scale shift, and higher-order shape shift, respectively. We set $\eta_{ik}=a_kX_{ik}$, $k=1,2,3$, with $(a_1,a_2,a_3)=(0.045,0.025,0.0022)$.
Since \(\E(\eta_{ik})=0\), it follows that \(\E\{Y_i^{-1}(u)\}=\mu^{-1}_{\oplus}(u)\). 

For the conditional generative model, we specify the biased and flattened surrogate barycenter
$\nu_{\oplus}^{-1}(u)
=\mu^{-1}_{\oplus}(0.5)+0.015+0.6\{\mu^{-1}_{\oplus}(u)-\mu^{-1}_{\oplus}(0.5)\}$. Surrogate informativeness is controlled through the latent coefficients $\zeta_{ik}
=
s_kd_k\eta_{ik}
+
\sqrt{1-d_k^2}\,\eta_{ik}',$
where \(\eta_i'\) is an independent copy of \(\eta_i\), \(s_k\in\{-1,1\}\) determines the alignment direction, and \(d_k\in[0,1]\) controls its alignment strength. Given \(X_i\), the fixed surrogate quantile function is set as
$G(X_i)^{-1}(u)
=
\nu^{-1}_{\oplus}(u)
+
\sum_{k=1}^{3}\zeta_{ik}\phi_k(u).$

We consider four sign regimes. Specifically, $\texttt{+++}$ corresponds to $(s_1,s_2,s_3)=(1,1,1)$, $\texttt{+-+}$ to $(1,-1,1)$, $\texttt{-+-}$ to $(-1,1,-1)$, and $\texttt{---}$ to $(-1,-1,-1)$. The first and last regimes correspond to fully positively and negatively aligned surrogate generators, respectively, while $\texttt{+-+}$ and $\texttt{-+-}$ represent mixed-alignment settings in which different functional directions require calibration coefficients with different signs. We consider three alignment strengths: $\texttt{strong}: d=(0.95,0.8,0.7)$, $\texttt{medium}: d=(0.8,0.6,0.5)$, and $\texttt{weak}: d=(0.3,0.1,0.1)$. These parameters ensure that \(Y_i^{-1}\) and \(G(X_i)^{-1}\) are valid quantile functions. For each labeled distribution, we observe $Y_{ij}=Y_i^{-1}(U_{ij}), j=1,\ldots,m,$ where \(U_{ij}\sim\mathrm{Unif}(0,1)\). As conditional generative models can produce arbitrarily many samples, we treat \(G(X_i)^{-1}\) as known.

\begin{comment}

\begin{figure}
    \centering
    \includegraphics[width=0.85\linewidth]{figures/motif/K562_dgp_comparison_DLEU1.png}
    \caption{Observed and generated quantile curves in real and simulated data. Left: observed K562 and State-generated quantile curves for the selected response gene.
Right: analogous simulated observed and generated quantile curves under the specified surrogate alignment and strength.
Each panel shows two representative sample-level curves and the corresponding barycenter curve.
}
    \label{fig:k562-dgp}
\end{figure}
\end{comment}
Under this construction, the oracle borrowing operator has an explicit form. Let $\Phi(u)=\{\phi_1(u),\phi_2(u),\phi_3(u)\}^{\top}.$
For any centered surrogate function satisfying $f(u)-\nu_{\oplus}(u)=\Phi(u)^\top v,$
the oracle operator acts as
$(\Gamma_*f)(u)=\Phi(u)^\top B_*v,
$
where
$B_*
=(1+\rho)^{-1}
\operatorname{diag}(s_1d_1,$ $s_2d_2,\ s_3d_3).$
Equivalently, $\Gamma_*(s,u)
=(1+\rho)^{-1}
\sum_{k=1}^3s_kd_k\phi_k(s)\phi_k(u).$
This expression shows that mixed-alignment regimes such as \(\texttt{+-+}\) and \(\texttt{-+-}\) require direction-specific calibration signs, which cannot be represented by scalar tuning.

We use two reference distributions. The uniform reference satisfies
\(\lambda_{\mathrm{unif}}^{-1}(u)=u\). For the Wasserstein test, we construct a control reference indexed by \(\delta\). Let
\(T_\delta(t)=t-\delta\sin(\pi t)\), and define
$\lambda_{\mathrm{ctrl},\delta}^{-1}(u)
=
T_\delta^{-1}\{\mu_\oplus^{-1}(u)\}.$
By construction, \(T_\delta\) is the optimal transport map from
\(\lambda_{\mathrm{ctrl},\delta}\) to \(\mu_\oplus\), with
\(\delta=0\) corresponding to the null and \(\delta>0\) to alternatives with increasing displacement.

%\paragraph{Comparison methods}
We compare four methods: \texttt{Labeled}, which uses labeled data only; \texttt{Naive GPI}, with $\Gamma=(1+\rho)^{-1}I$; \texttt{Scalar GPI}, with $\Gamma=\gamma I$; and \texttt{General GPI}, which uses the operator proposed in Section~\ref{subsec:Gamma-estimation}. Each setting is replicated 1000 times, with $B=1000$ bootstrap samples used for SCBs and Wasserstein tests. All quantile functions are interpolated onto a common grid of 199 equally spaced quantile levels over \([0.01,0.99]\). See more implementation details in
Supplementary Section~\ref{sec:implement-detail}.

%\subsection{Results for Different Estimation and Inference Tasks}
%\label{subsec:simulation-results}

\subsection{Estimation and Inference for Barycenter quantile function and linear displacement functionals}\label{subsec:simu-qf}
\label{subsubsec:barycenter-quantile}
\begin{figure}[t]
    \centering
    \includegraphics[width=\linewidth, height=0.38\textheight]{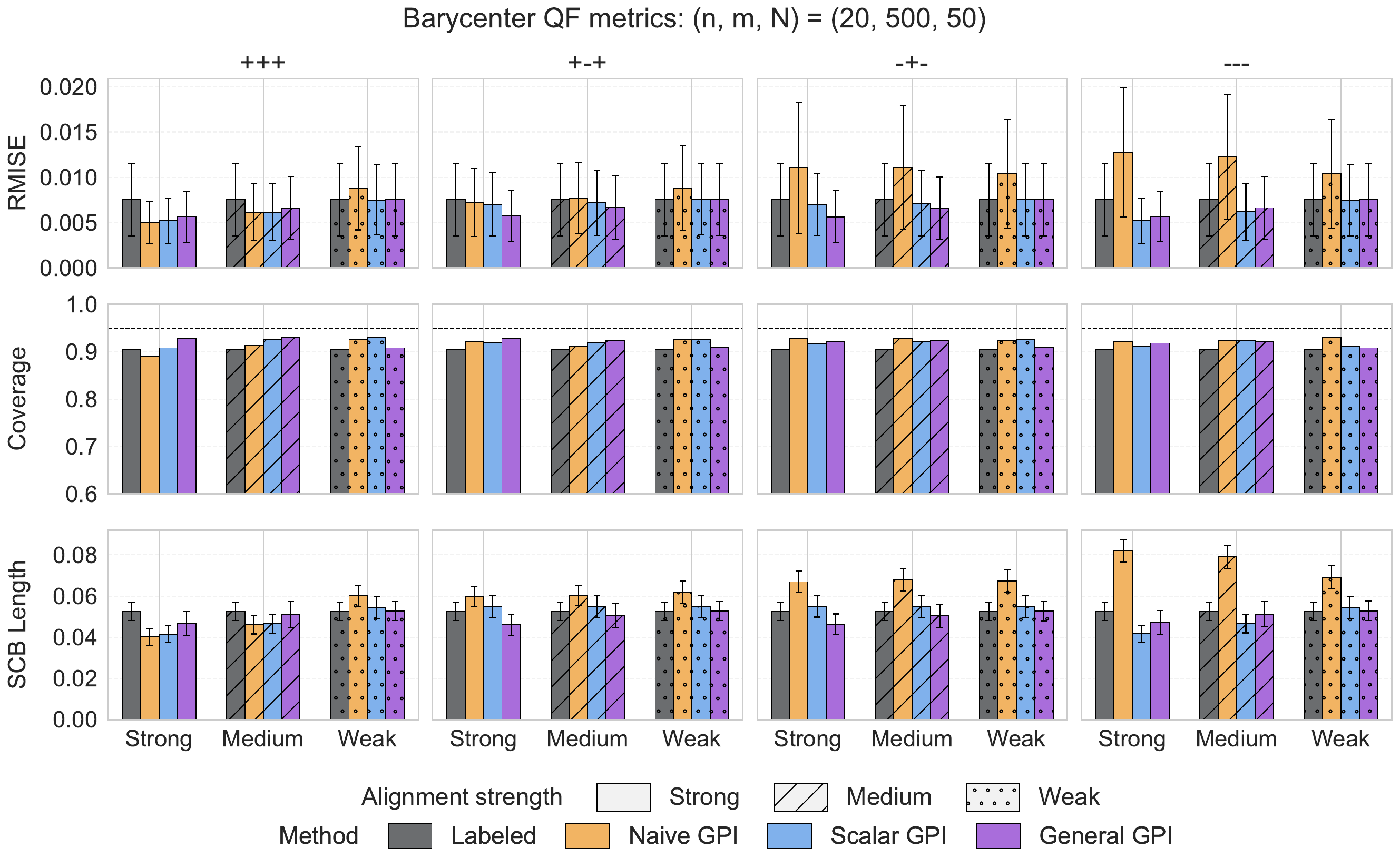}
    \captionsetup{skip=2pt,font={stretch=0.9}}
    \caption{\small Estimation and simultaneous inference for the barycenter quantile function across generative-surrogate alignment regimes under $(n,m,N)=(20,500,50)$.}
    \label{fig:qf}
\end{figure}

Under the twelve surrogate settings described in Section~\ref{subsec:simu-setup}, each method is used to estimate the quantile function of barycenter $\mu_\oplus$ and constructs a nominal 95\% SCB as in \eqref{eq:scb}. Figure~\ref{fig:qf} summarizes the root mean integrated squared error (RMISE), empirical SCB coverage, and average band length.

Empirical SCB coverage is comparable across methods and remains near the nominal level in most regimes, supporting the validity of the covariance estimation and inference procedure. The main differences among methods are reflected in RMISE and band length. Scalar and general GPI perform at least as well as the labeled-only method in nearly all regimes, with larger gains under stronger alignment. By contrast, naive GPI can yield larger RMISE and longer bands when the dominant functional direction is negatively aligned, as in the $\texttt{-+-}$ and $\texttt{---}$ regimes, highlighting the necessity of tuning. In the mixed-alignment regimes $(\texttt{+-+}/\texttt{-+-})$, general GPI benefits from direction-specific calibration and typically outperforms scalar GPI, which may even produce slightly longer bands than the labeled-only method. Nevertheless, scalar GPI remains more robust than naive GPI because its data-driven tuning is largely governed by the dominant first direction. In the homogeneous-alignment regimes $(\texttt{+++}/\texttt{---})$, scalar GPI can slightly outperform general GPI because a global calibration coefficient is sufficient. Overall, general GPI is the most robust across regimes, providing competitive improvements in RMISE and shorter band length while maintaining valid coverage. 

%Supplementary Section~\ref{sec:add-simu} reports additional simulations for the displacement functionals and alternative $(n,m,N)$ settings, with results consistent with the main findings.

\subsection{Wasserstein Testing for Barycenter-Control Comparison}
\label{subsec:simu-w2}
\begin{figure}[t]
    \centering
    \includegraphics[width=0.9\linewidth]{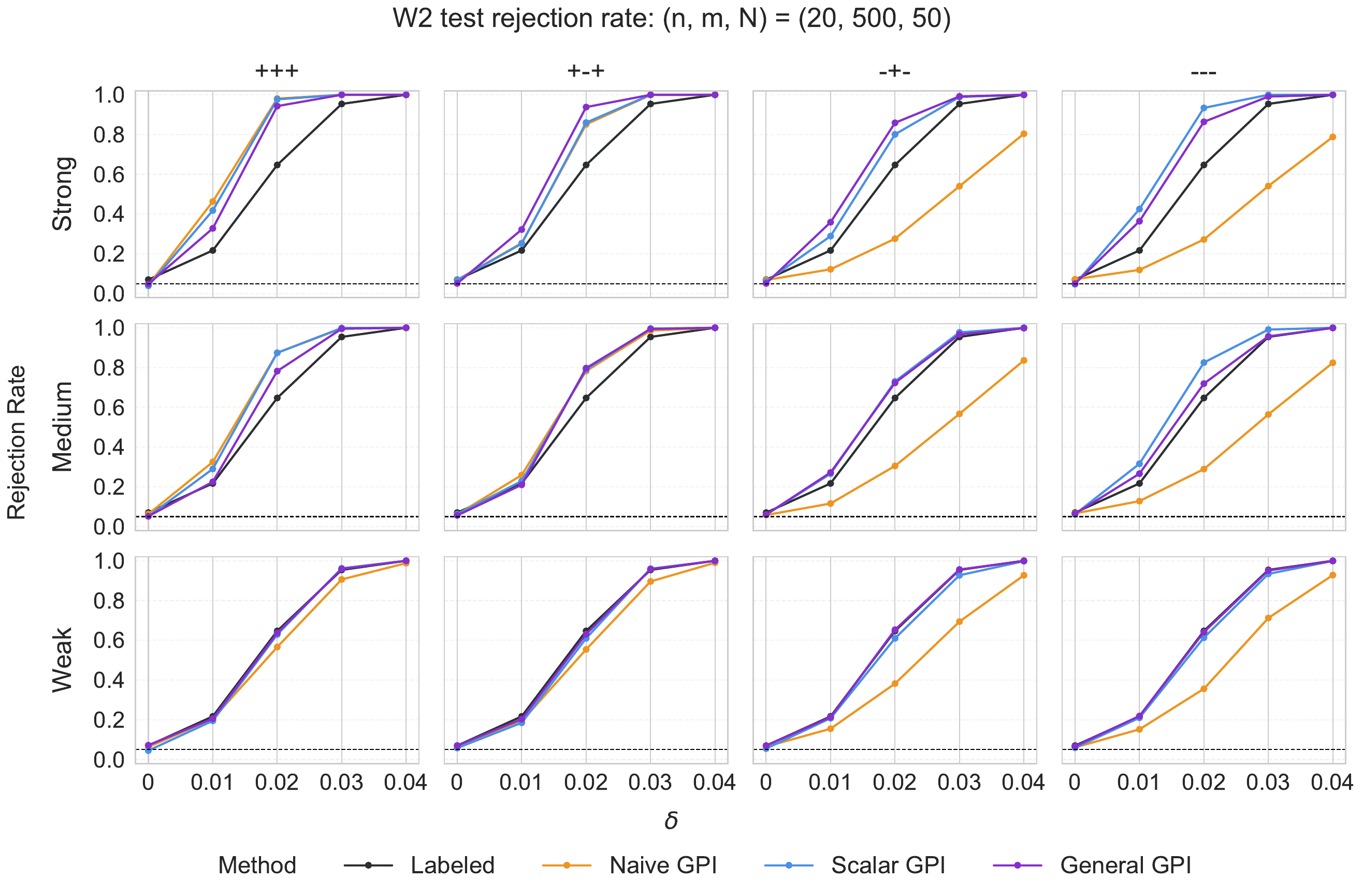}
    \captionsetup{skip=2pt,font={stretch=0.9}}
    \caption{\small Rejection rates of the Wasserstein test across surrogate-alignment regimes and displacement levels \(\delta\) under $(n,m,N)=(20,500,50)$. }
    \label{fig:w2-test}
\end{figure}

We next evaluate the Wasserstein test in Section~\ref{subsec:w2-test} for comparing the barycenter distribution with a control distribution used as the bridge reference. For each \(\delta\in\{0,0.01,0.02,0.03,0.04\}\), we set \(\lambda=\mu_0=\lambda_{\mathrm{ctrl},\delta}\) as defined in Section~\ref{subsec:simu-setup}. Under each surrogate alignment regime, we report the empirical rejection rates at the nominal level \(\alpha=0.05\). The results are summarized in Figure~\ref{fig:w2-test}.

Under the null, all methods maintain rejection rates near the nominal level across alignment regimes, supporting the validity of the proposed test. Under alternatives, scalar and general GPI achieve power at least comparable to the labeled-only test in nearly all regimes, with larger gains under stronger alignment. Naive GPI is more sensitive to the alignment pattern, improving power under positive alignment, as in the \(\texttt{+++}\) and \(\texttt{+-+}\) regimes, but potentially underperforming the labeled-only test under weak or negative alignment, particularly in the \(\texttt{-+-}\) and \(\texttt{---}\) regimes. General GPI performs better in mixed-alignment regimes, especially under strong alignment, whereas scalar GPI can slightly outperform it in homogeneous-alignment regimes such as $\texttt{+++}/\texttt{---}$, where a single global calibration coefficient is sufficient. Overall, general GPI provides the most robust performance, maintaining valid size control and competitive power, consistent with the patterns observed in Section~\ref{subsec:simu-qf}.

\subsection{Additional simulation results} 
Additional simulations in the Supplementary Section \ref{sec:add-simu} further support these findings. First, for linear displacement functionals, including mean, central-region, and upper-tail effects, General GPI consistently provides efficiency gains while maintaining approximately nominal coverage across diverse surrogate-alignment regimes; in contrast, Naive GPI can lose efficiency under negative or heterogeneous alignment (Figures~\ref{fig:tail-displacement}-\ref{fig:mean}).  Second, as the unlabeled sample size increases, General GPI increasingly benefits from the additional surrogate information, with lower estimation error, shorter confidence regions, and increased testing power while maintaining valid coverage and type-I error control (Figure~\ref{fig:metric_vs_N}).  Finally, simulations with both smaller and larger sample sizes show qualitatively similar patterns: GPI continues to improve estimation and inference in smaller samples, while its efficiency gains—and the robustness advantage of General GPI under mixed or weak alignment—become more pronounced with larger samples (Figures~\ref{fig:qf-small-m}–\ref{fig:w2-test-large-n-m}).
\section{Application to K562 perturb-seq data}\label{sec:real}
\subsection{Data, estimand, and generative surrogate}

%In large-scale Perturb-seq experiments, perturbations with similar biological functions are often grouped according to pathway annotations or shared transcriptional response patterns. For each perturbed gene, the outcome of interest is naturally represented by the distribution of expression levels of a target gene across all sequenced cells. These perturbation-specific expression distributions characterize how the target gene responds to different but functionally related genetic perturbations. A biologically meaningful summary of the collective perturbation effect is the Wasserstein barycenter of these expression distributions, which represents the average transcriptional response of the perturbation group. However, the number of observed cells can vary substantially across perturbations because of differences in perturbation efficiency, cell viability, and sequencing depth, resulting in heterogeneous estimation precision across perturbation-specific distributions and, consequently, the group-level barycenter.

In large-scale Perturb-seq experiments, functionally related perturbations are often grouped using pathway annotations or shared transcriptional response patterns. For a given response gene, each perturbation induces an expression distribution across sequenced cells, and their Wasserstein barycenter provides a natural summary of the average response within the group. Uneven cell coverage can leave some perturbation-specific distributions poorly measured or unavailable, motivating the use of generative surrogates for barycenter estimation.

We illustrate our framework using the K562 Perturb-seq dataset of \citet{replogle2022mapping}, which combines CRISPR interference with single-cell RNA sequencing to characterize transcriptional responses to large-scale genetic perturbations.  We focus on the 40S ribosome (40Sr), a coherent functional module involved in small ribosomal subunit assembly and function. Perturbations of individual 40Sr genes are expected to induce a shared transcriptional program related to impaired ribosome biogenesis and translational stress while preserving gene-specific heterogeneity. Using Gene Ontology annotations, we identify 160 perturbation targets related to 40Sr.

To generate synthetic perturbation responses, we use State
\citep{adduri2025predicting}, a transformer-based foundation model that generates virtual post-perturbation cell populations from control cells and perturbation labels. State was pretrained on approximately 167 million single-cell transcriptomes and fine-tuned on perturbation data from non-K562 cell lines, with K562 perturbations excluded from both stages. After quality control, 74 perturbations remain in the 40Sr cluster. We treat the 21 perturbations with more than 200 observed cells as labeled and the remaining 53 as unlabeled, retaining their observed data only for benchmarking. Control cells serve as input to State. Further details are provided in Supplementary Section~\ref{subsec:app-detail}.

For a given response gene, let $\mu_\oplus$ denote the population Wasserstein barycenter of the expression distributions induced by perturbations within the 40Sr module. Biologically, $\mu_\oplus$ represents the consensus transcriptional response of that gene to perturbations of the pathway. For perturbation $i$, let $Y_i$ denote its gene-expression distribution across sequenced cells, and $G(X_i)$ the surrogate distribution generated by State, where $X_i$ identifies the perturbation. We use the uniform distribution on $[0,1]$ as the bridge reference. Let \(\mu_0\) denote the expression distribution of the same response gene among the 10,691 K562 non-targeting control cells, which is treated as effectively known. We characterize the pathway-level perturbation effect through the quantile displacement $\mu_\oplus^{-1}(u)-\mu_0^{-1}(u), u\in[0,1],$ which quantifies how perturbations of the 40Sr pathway shift the expression distribution relative to the control across quantile levels. We further summarize these effects using the mean (\(A=[0,1]\)), central-region (\(A=[0.25,0.75]\)), and upper-tail (\(A=[0.9,1]\)) functionals, together with the Wasserstein distance \(W_2(\mu_\oplus,\mu_0)\).
 
Since the population Wasserstein barycenter is unknown, we use the empirical barycenter based on all 74 observed perturbation distributions as a complete-coverage benchmark (\texttt{Full observed}). We also consider a hybrid benchmark (\texttt{Pooled}) that combines the 21 labeled distributions with State-generated surrogates for the remaining 53 perturbations treated as unlabeled. Uncertainty for both benchmarks is estimated using 1,000 two-stage bootstrap replicates that resample perturbations and cells, with surrogate distributions treated as fixed for \texttt{Pooled}.

%Since the population Wasserstein barycenter is unknown, we use the empirical barycenter computed from all 74 observed perturbation distributions as the benchmark (\texttt{Full observed}), representing the estimator that would be available if all perturbations were experimentally measured. Uncertainty is quantified using a two-stage bootstrap that resamples perturbations and, within each perturbation, resamples cells to account for both between- and within-perturbation variability.

%As a second benchmark, we consider a hybrid estimator (\texttt{Pooled}) that combines the observed distributions from the 21 labeled perturbations with the State-generated surrogate distributions for the remaining 53 unlabeled perturbations. Its uncertainty is estimated using an analogous bootstrap procedure, treating the surrogate distributions as fixed. Both benchmark procedures use 1,000 bootstrap replicates.

\subsection{Empirical validation of GPI calibration}
\label{subsec:appl-gene-level-analysis}
To examine how surrogate alignment affects GPI, we classify response genes using the labeled perturbations. For each gene, we project the observed and State-generated distributions onto the location, scale, and shape directions from Section~\ref{subsec:simu-setup}. The signs of the three observed--surrogate score correlations determine the alignment pattern, and their average absolute magnitude determines the alignment strength. After preprocessing, 1,980 response genes remain; details are provided in Supplementary Section~\ref{subsec:app-detail}.

\begin{figure}[t]
    \centering
    \includegraphics[width=\linewidth]{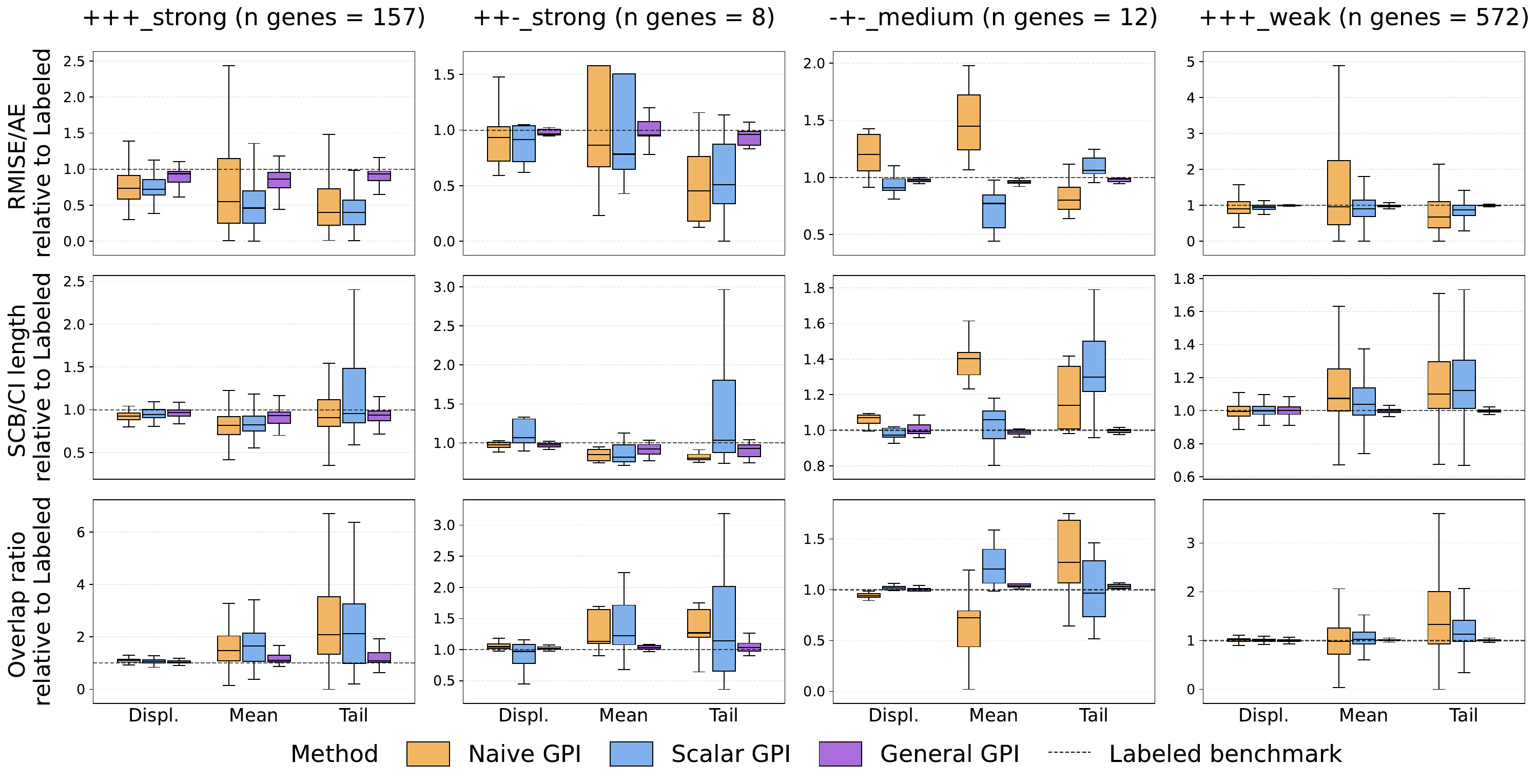}
    \captionsetup{skip=2pt,font={footnotesize,stretch=0.9}} 
    \caption{\small Performance ratios of GPI methods relative to \texttt{Labeled} across alignment patterns, based on estimation error, confidence-region width, and overlap with \texttt{Full observed} for the displacement function (Displ.), mean, and upper-tail displacement.}
    \label{fig:all_genes_metric_ratio}
\end{figure}

\begin{figure}[t]
    \centering
    \includegraphics[width=\linewidth]{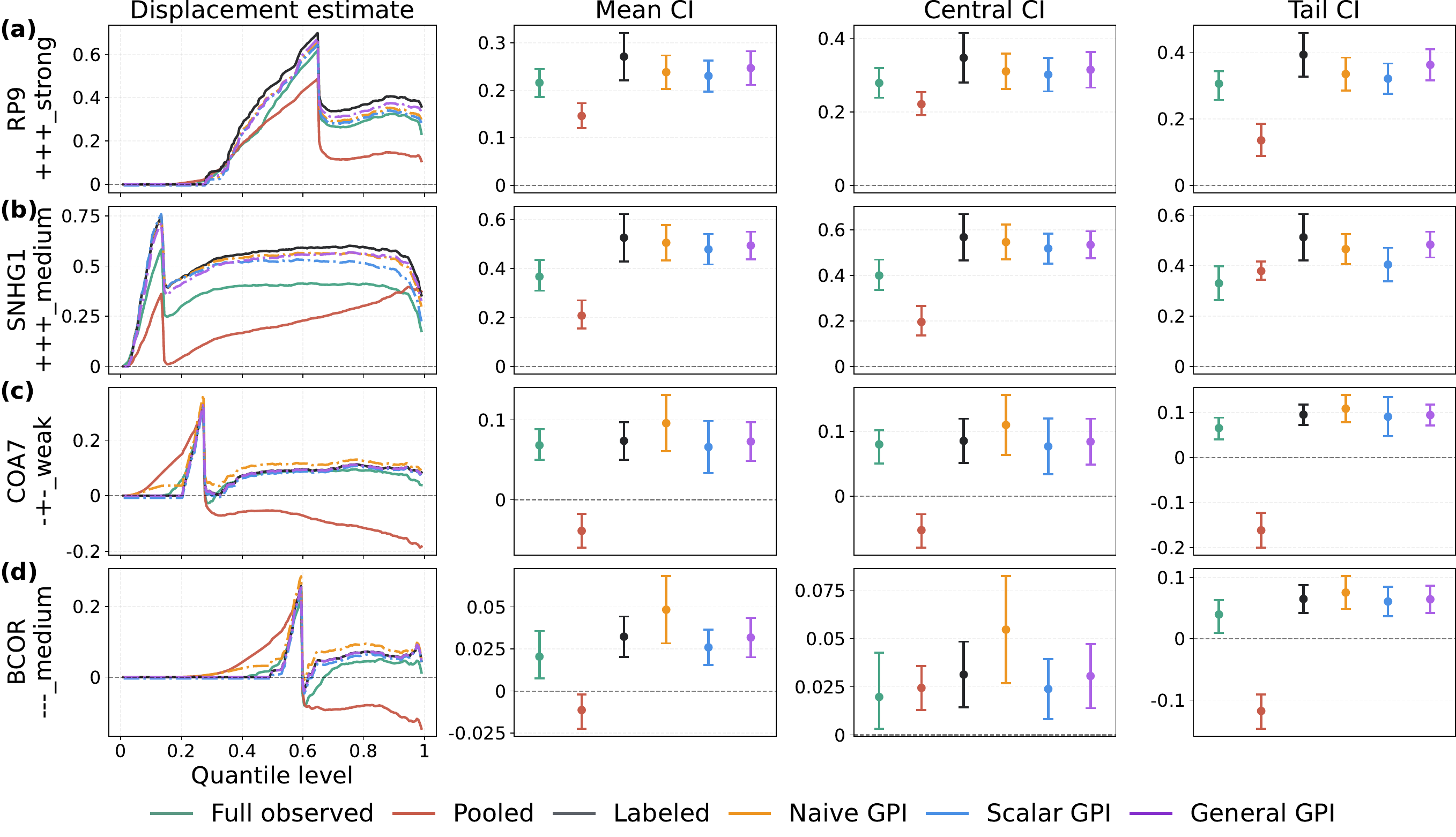}
    \captionsetup{skip=2pt,font={footnotesize,stretch=0.9}}
    \caption{\small Quantile displacement estimates and inference for representative genes across surrogate-alignment patterns, including mean, central-region, and upper-tail functionals.}
    \label{fig:pattern_representative_genes_summary}
\end{figure}

We focus on four representative groups: strong \texttt{+++}, strong \texttt{++-}, medium \texttt{-+-}, and weak \texttt{+++}. Using \texttt{Full observed} as the benchmark, we evaluate displacement-function RMISE, absolute errors (AE) of displacement functionals, SCB and CI widths, and overlap with the \texttt{Full observed} confidence regions, where the overlap ratio is the intersection area or length divided by that of the evaluated confidence region. Figure~\ref{fig:all_genes_metric_ratio} reports these metrics relative to \texttt{Labeled}. Under strong alignment, all three GPI methods improve over \texttt{Labeled}, with larger gains under \texttt{+++} than \texttt{++-}. \texttt{Naive GPI} and \texttt{Scalar GPI} often achieve larger gains but are more variable, whereas \texttt{General GPI} is more conservative and stable. Its advantage is clearest under mixed or weak alignment, where it avoids substantial losses from unreliable surrogate information. 
Supplementary Figure S12 further compares \texttt{Pooled}, which often yields narrower confidence regions but substantially larger estimation errors and poorer agreement with the \texttt{Full observed} benchmark, particularly under mixed or weak alignment. This highlights the importance of calibrating, rather than directly substituting, generative-model outputs for observed data. Figure~\ref{fig:pattern_representative_genes_summary} illustrates these patterns using RP9 (strong \texttt{+++}), SNHG1 (medium \texttt{+++}), COA7 (weak \texttt{-+-}), and BCOR (medium \texttt{---}). Additional gene-level results are provided in Supplementary Section~\ref{subsec:add-app}.

\begin{figure}[t]
    \centering
    \includegraphics[width=\linewidth]{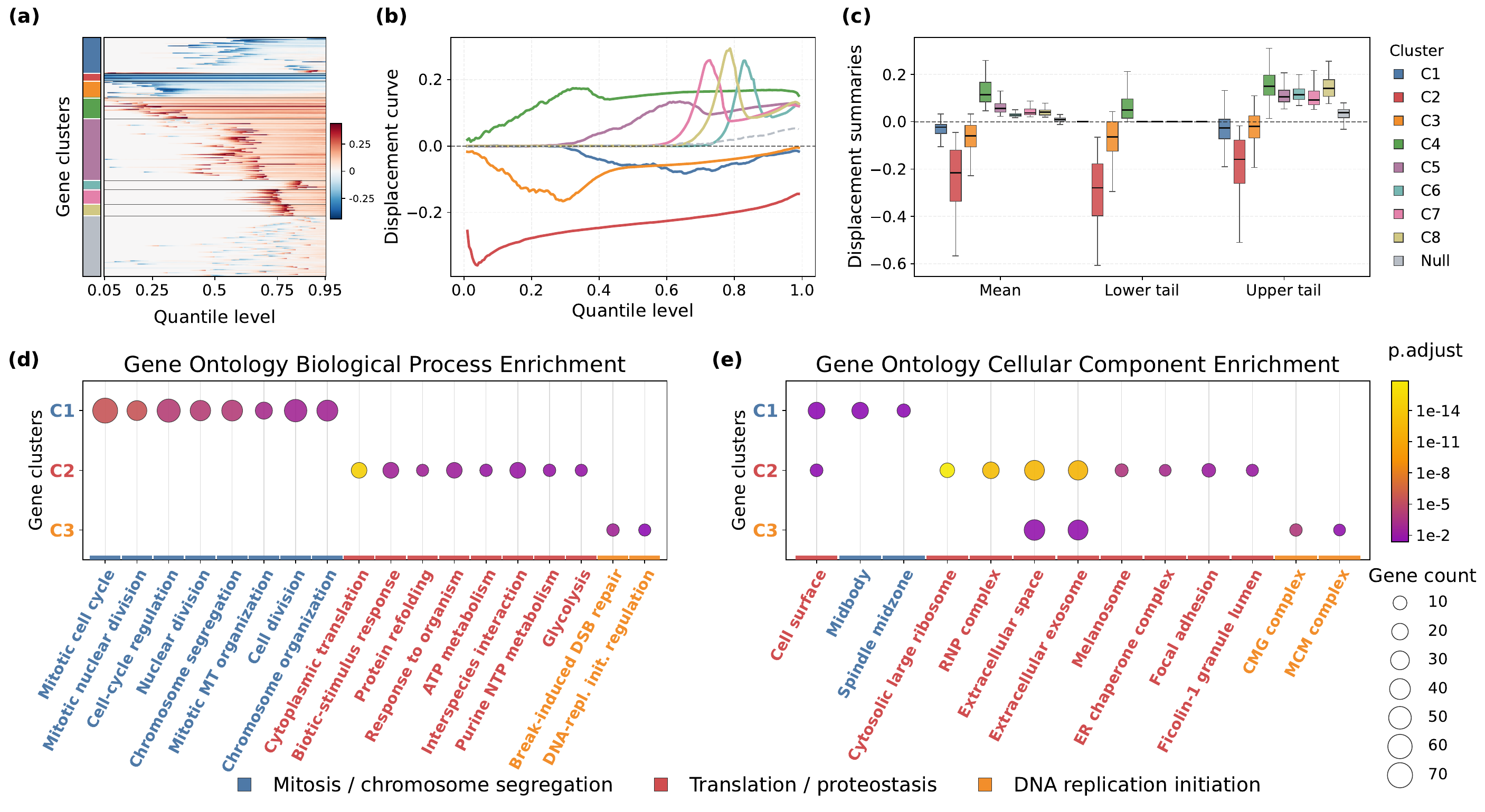}
    \caption{\small
Quantile-level displacement patterns and associated biological programs.
(a) Gene-wise displacement heatmap by \texttt{General GPI}.
(b) Cluster-average displacement curves.
(c) Mean, lower-tail, and upper-tail summaries.
(d) GO Biological Process enrichment.
(e) GO Cellular Component enrichment.}
    \label{fig:GO_enrichment}
\end{figure}
\subsection{Landscape of distributional perturbation effects}
The displacement estimates reveal substantial across-quantile heterogeneity among the 1,980 response genes, with many effects localized beyond average shifts. Gene-wise displacement heatmaps further show that the \texttt{General GPI} estimates more closely reproduce the \texttt{Full observed} displacement patterns than direct pooling of observed and State-generated distributions (Supplementary Figure~\ref{fig:displacement-heatmap}), providing evidence that calibration removes systematic surrogate errors while retaining useful distributional information. To assess whether this heterogeneity reflects biological structure, we cluster genes using FPC scores from the \texttt{General GPI} displacement estimates and perform Gene Ontology enrichment analysis. The resulting clusters exhibit distinct quantile-level response patterns with coherent biological enrichment (Figure~\ref{fig:GO_enrichment}). Clustering details are provided in Supplementary Section~\ref{subsec:add-app}.

%To further interpret these structured distributional responses, we apply FPCA to the \texttt{Full observed} displacement curves and cluster genes based on their leading FPCA scores, followed by functional enrichment analysis within the resulting displacement-profile clusters (Figure~\ref{fig:GO_enrichment}a--c). This analysis shows that distinct quantile-level response patterns correspond to coherent biological programs, including mitotic chromosome segregation, translation/proteostasis, and DNA-replication-related processes, as supported by their enriched Gene Ontology Biological Process and Cellular Component terms (Figure~\ref{fig:GO_enrichment}d--e).

The enriched clusters partially agree with those reported in \citet{replogle2022mapping}, while displacement-based clustering offers a finer view of quantile-dependent responses that may reflect cell-state heterogeneity. For cell-cycle-related programs, expression quantiles may partially reflect state activity. C1, enriched for mitotic chromosome-segregation processes, shows middle-to-upper quantile attenuation, consistent with suppression of the mitotic high-expression tail. C2, enriched for translation and proteostasis, exhibits broadly negative displacement, consistent with widespread effects of 40S ribosome perturbation. C3, enriched for DNA-replication initiation, shows lower-to-middle quantile suppression with a relatively preserved upper tail, suggesting stronger effects on less replication-active cells.

\section{Discussion}\label{sec:Discuss}
We have proposed a general framework for generation-powered inference (GPI) for distribution-valued outcomes. Using a bridge representation based on transformed quantile functions, GPI converts inference on nonlinear Wasserstein barycenters into inference on function-valued parameters in a linear Hilbert space. This enables principled integration of AI-generated and experimentally observed distributions while preserving valid inference under imperfect generative models, extending prediction-powered inference beyond finite-dimensional Euclidean parameters.

More broadly, GPI provides a framework for using generative models as auxiliary information rather than direct substitutes for observations, with their systematic biases calibrated by labeled data. Several extensions are of interest, including settings with covariate shift between labeled and unlabeled populations \citep{ying2025towards} and the integration of multiple generative models with heterogeneous predictive quality \citep{shan2025sada}. We hope that this work stimulates further research on principled statistical inference with AI-generated surrogate distributions.

%We have proposed a general framework for generation-powered inference (GPI) for distribution-valued outcomes. By introducing a bridge representation based on transformed quantile functions, we convert statistical inference on nonlinear Wasserstein barycenters into inference on function-valued parameters in a linear Hilbert space. This bridge representation enables principled integration of AI-generated and experimentally observed distributions while preserving valid inference under imperfect generative models. As a result, GPI substantially extends the scope of prediction-powered inference from finite-dimensional Euclidean parameters to a broad class of distribution-valued inferential problems.

%More broadly, GPI offers a new paradigm for statistical inference with generative models. Rather than treating AI-generated data as direct substitutes for observations, GPI uses them as auxiliary information  calibrated by experimentally observed samples. This perspective shows that reliable statistical inference remains possible even when generative models are imperfect, provided that their systematic biases can be learned from labeled data. We hope that this work stimulates further research on principled statistical inference with AI-generated surrogate data.

\section*{Data availability}
The original K562 Perturb-seq data are available at \url{https://gwps.wi.mit.edu}. The pretrained State model used in this study are available at \url{https://huggingface.co/arcinstitute/ST-SE-Replogle/tree/main/zeroshot/k562}. 

\section*{Acknowledgments}
This research is supported by NIH grants R01GM129781 and U01HG013841.		
\printbibliography
\end{document}